\pdfoutput=1

\PassOptionsToPackage{hyphens}{url}
\documentclass[camera,letterpaper,nomarginnotes,nonarrowgutter]{jpaper}

\usepackage{siunitx}

\usepackage[noadjust]{cite}
\usepackage{amsmath,amssymb,amsfonts}
\usepackage{algorithmic}
\usepackage{graphicx}
\usepackage{soul}
\usepackage{textcomp}
\usepackage[dvipsnames]{xcolor}
\usepackage{balance}
\usepackage{mathptmx} 
\usepackage{fancyhdr}
\usepackage[normalem]{ulem}
\usepackage{footmisc}

\usepackage{marginnote}
\usepackage{booktabs} 
\usepackage{xspace}
\usepackage{enumitem}
\usepackage{longfbox}

 \usepackage{amsmath}
 \usepackage{ifthen}
 \usepackage{multirow}
 \usepackage{listings}
 \usepackage{setspace}
 \usepackage{float}
 \usepackage[us,12hr]{datetime}
 \usepackage{tikz}

\usepackage[colorinlistoftodos,prependcaption,textsize=small]{todonotes}
\usepackage{xargs}                    
\usepackage{titletoc}
\usepackage{clipboard}
\usepackage[all]{nowidow}
\usepackage[framemethod=tikz]{mdframed}
\usepackage[most]{tcolorbox}

\tcbuselibrary{breakable}
\tcbuselibrary{hooks}
\usepackage[keeplastbox]{flushend}
\usepackage[bookmarks=true,breaklinks=true,letterpaper=true,colorlinks,linkcolor=black,citecolor=blue,urlcolor=black]{hyperref}

\usepackage{cleveref}

\crefformat{section}{\S#2#1#3}
\crefformat{subsection}{\S#2#1#3}
\crefformat{subsubsection}{\S#2#1#3}

\newcommand{\gfcr}[1]{\textcolor{black}{#1}} 
\newcommand{\gfcri}[1]{\textcolor{black}{#1}} 
\newcommand{\sgcri}[1]{\textcolor{black}{#1}} 
\newcommand{\gfcrii}[1]{\textcolor{black}{#1}} 
\newcommand{\gfcriii}[1]{\textcolor{black}{#1}}
    \newcommand{\sgcriii}[1]{\textcolor{black}{#1}}
    \newcommand{\gfcriv}[1]{\textcolor{black}{#1}}
    \newcommand{\prtag}[1]{}

    \newcommandx{\unsure}[2][1=]{}
    \newcommandx{\changev}[2][1=]{}
    \newcommandx{\feedback}[2][1=]{}
    \newcommandx{\improvement}[2][1=]{}
    \newcommandx{\thiswillnotshow}[2][1=]{}
    \newcommandx{\completedRevision}[2][1=]{}
    \newcommandx{\dataSource}[2][1=]{}
    \newcommandx{\info}[2][1=]{}

\newcommand{\paperlinespacing}[1]{0.7}

\newcommand{\change}[1]{{#1}}
\newcommand{\paratitle}[1]{\vspace{4pt}\noindent\textbf{#1.}}
\newcommand{\titleShort}{Polynesia\xspace}
\newcommand{\sgdel}[1]{}
\newcommand{\sgmod}[2]{#2}
\newcommand{\am}[1]{#1}
\newcommand{\sgii}[1]{{#1}}
\newcommand{\sgiii}[1]{{#1}}
\newcommand{\sgiv}[1]{{#1}}

\newcommand{\rev}[1]{{#1}}
\newcommand{\gf}[1]{\textcolor{black}{#1}} 
\newcommand{\sg}[1]{\textcolor{black}{#1}} 
\newcommand{\gfdel}[1]{}

\newcommand{\gfrev}[1]{\textcolor{black}{#1}} 
\newcommand{\sgrev}[1]{\textcolor{black}{#1}} 
\newcommand{\sgrevi}[1]{\textcolor{black}{#1}}

\newcommand{\incircle}[1]{%
  \tikz[baseline=(char.base)]{
  \node[shape=circle,draw,inner sep=0.5pt,fill=black,text=white,font=\bfseries\footnotesize] (char) {#1};}%
}%
\newcommand{\incircledd}[1]{%
  \tikz[baseline=(char.base)]{
  \node[shape=circle,draw,inner sep=0.5pt,fill=white,text=black,font=\bfseries\itshape\footnotesize] (char) {#1};}%
}%

\newcommand{\incircleddd}[1]{%
  \tikz[baseline=(char.base)]{
  \node[shape=circle,draw,inner sep=0.1pt,fill=black,text=white,font=\bfseries\footnotesize] (char) {#1};}%
}%

\titlespacing\section{2pt}{3pt plus 1pt minus 1pt}{2pt plus 1pt minus 1pt}
\titlespacing\subsection{2pt}{3pt plus 1pt minus 1pt}{2pt plus 1pt minus 1pt}
\titlespacing\subsubsection{2pt}{3pt plus 1pt minus 1pt}{2pt plus 1pt minus 1pt}

\makeatletter
\g@addto@macro{\normalsize}{%
  \setlength{\abovedisplayskip}{3pt plus 0.5pt minus 1pt}
  \setlength{\belowdisplayskip}{3pt plus 0.5pt minus 1pt}
  \setlength{\abovedisplayshortskip}{0pt}
  \setlength{\belowdisplayshortskip}{0pt}
  \setlength{\intextsep}{4pt plus 1pt minus 1pt}
  \setlength{\textfloatsep}{4pt plus 1pt minus 1pt}
  \setlength{\skip\footins}{5pt plus 1pt minus 1pt}}
\makeatother

\def\BibTeX{{\rm B\kern-.05em{\sc i\kern-.025em b}\kern-.08em
    T\kern-.1667em\lower.7ex\hbox{E}\kern-.125emX}}

\begin{document}

\bstctlcite{IEEEexample:BSTcontrol}

\title{\scalebox{0.91}{Enabling High-Performance and Energy-Efficient}\\ \scalebox{0.91}{Hybrid Transactional/Analytical Databases}\\ \scalebox{0.91}{with Hardware/Software Cooperation}}
\newcommand{\affilGoogle}[0]{$^\dagger$}
\newcommand{\affilUIUC}[0]{$^\diamond$}
\newcommand{\affilETH}[0]{$^\ddagger$}

\author{
Amirali Boroumand\affilGoogle%
\quad\qquad%
Saugata Ghose\affilUIUC%
\quad\qquad%
Geraldo F. Oliveira\affilETH%
\quad\qquad%
Onur Mutlu\affilETH\\
{\affilGoogle}\emph{Google}%
\qquad%
 {\affilUIUC}\emph{Univ.\ of Illinois Urbana-Champaign}%
 \qquad
 {\affilETH}\emph{ETH Z{\"u}rich}%
 \vspace{-5pt}
}

\maketitle
    
\thispagestyle{plain}
\pagestyle{plain}

\begin{abstract}

\sgiii{A growth in data volume, combined with increasing demand for real-time analysis (using the most recent data), has resulted in the emergence of database systems that concurrently support transactions and data analytics. These \emph{hybrid transactional and analytical processing} (HTAP) database systems can support real-time data analysis without the high costs of synchronizing across separate single-purpose databases.} \sgii{Unfortunately, for many applications that perform a high rate of data updates, state-of-the-art HTAP systems incur significant \gfcri{losses} in transactional (up to 74.6\%) and/or analytical (up to 49.8\%)
throughput compared to performing only transaction\gfcri{al} or only
analytic\gfcri{al queries} in isolation, due to
(1)~data movement between the CPU and memory,
(2)~data update propagation \gfcri{from transactional to analytical workloads}, and
(3)~\gfcri{the} cost \gfcri{to maintain a \sgcri{consistent} view of data across the system}.}

\Copy{CR1/1A}{We propose \sgrev{\emph{\titleShort},} a hardware--software co-designed \rev{system for in-memory HTAP databases} \gfcri{that avoids the large throughput losses of traditional HTAP systems}. \sgii{\titleShort
(1)~divides the HTAP system into \am{transactional and analytical processing islands,}
(2)~\sgiv{implements} \sgrev{new custom hardware that unlocks software optimizations} 
to reduce the costs of update propagation and consistency, and
(3)~\am{exploits} processing-in-memory for the analytical islands to alleviate data movement \gfcri{overheads}.}
}\sgii{Our evaluation shows that \titleShort outperforms three state-of-the-art
HTAP systems, with average transactional/analytical throughput improvements of
1.7$\times$/3.7$\times$, and reduces
energy consumption by 48\% over the prior lowest-energy \gfcri{HTAP} system.}
\end{abstract}

\section{Introduction}
\label{sec:intro}

Data analytics has become popular due to the \gfcr{rapid} growth
\gfcri{of} data generated annually~\cite{cisco-report}.
Many application domains\gfcr{, such as fraud detection~\cite{cao2019titant,qiu2018real,quah2008real}, business intelligence~\cite{sql-htap,snappy-data,sahay2008real}, healthcare~\cite{chisholm2014adopting,ta2016big}, personalized recommendation~\cite{wiser,zhou2017kunpeng}, and IoT~\cite{wiser},} have a critical need to perform 
\emph{real-time data analysis}, \gfcri{where data analysis needs to be performed using the most recent version of data~\cite{real-time-analysis-sql,huang2020tidb}. To enable real-time data analysis, state-of-the-art database management systems (DBMSs) leverage \emph{hybrid transactional and analytical processing} (HTAP)~\cite{htap-gartner,sap-hana-evolution,htap}.} An HTAP DBMS is a single-DBMS
solution that supports both transactional and analytical
workloads~\cite{htap-gartner,peloton,batchdb,htap-survey,real-time-analysis-sql}.
\gfcri{Ideally, an} HTAP system should have three properties~\cite{batchdb} \gfcr{to guarantee efficient execution of transactional and analytical workloads}.
First, it should ensure that both transactional and analytical workloads 
\am{benefit from their own workload-specific optimizations (e.g., algorithms, data structures)}.
Second, it should guarantee \gfrev{data freshness and data consistency} \am{(i.e., access to \sgiv{the} most recent version of data)} for analytical workloads
while \am{ensuring \sgiv{that} both transactional and analytical workload\gfcri{s} have} a consistent view of data across the system.
Third, it should ensure that the latency and throughput of \gfcri{both} the transactional \gfcri{workload} and
\gfcri{the} analytical workload are the same as if \gfcri{each of them} were run in isolation.

We extensively study state-of-the-art HTAP systems (\gfcri{\cref{sec:bkgd:motiv}}) 
and observe two key problems that prevent them from achieving all three
properties of an ideal HTAP system.
First, \rev{these} systems experience a drastic reduction in
transactional throughput (up to 74.6\%) and analytical throughput (up to 49.8\%) compared to \am{when \gfcr{transactional and analytical workloads run} in isolation.} \am{This is because \sgiv{the mechanisms used} to provide data freshness and consistency}
induce a
significant amount of \emph{data movement} between the CPU cores and
main memory.
Second, HTAP systems often fail to provide effective performance isolation.
 These systems suffer from severe performance interference because of the \sgii{high resource contention}
between transactional workloads and analytical workloads. \emph{Our goal} in this work is to develop an HTAP system that
overcomes these problems while achieving all three of the
desired HTAP properties, \am{with new architectural techniques}.

\Copy{CR1/1B}{{We} propose a \rev{novel system 
for in-memory HTAP databases} called 
\sgrev{\emph{\titleShort}}. \sgiv{The key insight behind \titleShort is to partition the computing resources into two isolated \gfcr{new custom} processing \emph{islands}: \emph{transactional islands} and \emph{analytical islands}.} \gfcri{An island is a hardware--software co-designed component specialized for \gfcrii{specific} types of queries.} Each island consists of (1)~a replica of data for a specific workload, 
(2)~an optimized execution engine \am{(i.e., the \sgiv{software} that executes queries)}, and
(3)~a set of hardware resources (e.g., computation units, memory)
that cater to the execution engine and its memory access patterns.

\gfcr{\titleShort meets all desired properties from a\gfcri{n} HTAP system in three ways. First, by employing processing islands, \titleShort enables workload-specific optimizations for both transactional and analytical workloads (\emph{first desired HTAP property}).} \sgrev{Second, we design new hardware accelerators to add specialized capabilities to
each island, which we exploit to optimize the performance of several key HTAP algorithms.} This includes \sgrev{new accelerators and modified algorithms} \gfcr{to 
propagate transactional updates to analytical islands (\gfcri{\cref{sec:proposal:update-propagation}}) \gfcr{and to maintain \gfcr{a consistent view of data across the system  (\gfcri{\cref{sec:proposal:consistency}}). \gfcr{Such new components ensure data freshness and data consistency in our HTAP system (\emph{second desired HTAP property}).}}}}}
\gfcr{Third, we tailor the design of transactional and analytical islands to fit the characteristics of transactional and analytical workloads. The transactional islands use dedicated CPU hardware resources (i.e., multicore CPUs and multi-level caches) to execute transactional workloads since transactional queries have cache-friendly access patterns~\gfcriv{\cite{conda,LazyPIM,amiraliphd}}. The analytical islands leverage processing-in-memory (PIM) techniques~\gfcri{\cite{mutlu2020modern,pim-survey,ghose.ibmjrd19}} due to the large data traffic analytical workloads produce. 
PIM systems~\gfcriv{\cite{
Kautz1969,%
stone1970logic,%
kogge1994execube,%
gokhale1995processing,%
patterson1997case,%
oskin1998active,%
Draper:2002:ADP:514191.514197,%
ambit,%
seshadri.arxiv16,%
seshadri2013rowclone,%
seshadri2018rowclone,%
angizi2019graphide,%
kim.hpca18,%
kim.hpca19,%
chang.hpca16,%
hajinazarsimdram,%
rezaei2020nom,%
wang2020figaro,%
syncron,%
fernandez2020natsa,%
alser2020accelerating,%
cali2020genasm,%
kim.arxiv17,%
kim.bmc18,%
pim-enabled,%
tesseract,%
google-pim,%
conda,%
amiraliphd,
LazyPIM,%
boroumand.arxiv17,%
singh2019napel,%
DBLP:conf/sigmod/BabarinsaI15,%
7056040,%
Mingyu:PACT,%
gao2016hrl,%
gu.isca16,%
guo2014wondp,%
hashemi.isca16,%
cont-runahead,%
tom,%
neurocube,%
kim.sc17,%
DBLP:conf/IEEEpact/LeeSK15,%
liu-spaa17,%
morad.taco15,%
graphpim,%
pattnaik.pact16,%
pugsley2014ndc,%
toppim,%
zhu2013accelerating,%
data-reorganization-pim,%
tetris,%
mondrian,%
dai2018graphh,%
graphp,%
huang2020heterogeneous,%
zhuo2019graphq,%
santos2017operand,%
mutlu2020modern,%
ghose.ibmjrd19,%
ghose2019arxiv,%
besta2021sisa,
ferreira2021pluto,%
olgun2021quactrng,%
lloyd2015memory,%
elliott1999computational,%
landgraf2021combining,%
rodrigues2016scattergather,%
lloyd2017keyvalue,%
gokhale2015rearr,%
nair2015active,%
PIM_fall2021, %
olgun2021pidram, %
lee2022isscc, %
ke2021near,%
kwon202125, %
lee2021hardware, %
herruzo2021enabling, %
singh2021fpga, %
singh2021accelerating, %
oliveira2021pimbench,%
boroumand2021google,
boroumand2021google_arxiv,
denzler2021casper,
ghiasi2022genstore,
NIM} mitigate data movement bottlenecks by placing computation units nearby or inside memory\gfdel{ (e.g., SRAM caches~\cite{gao2020computedram,eckert2018neural,fujiki2019duality,kang.icassp14,denzler2021casper}, DRAM~\gfcriv{\cite{
tesseract, 
google-pim,
LazyPIM, 
conda, 
amiraliphd,
tom,
graphp,
tetris,
graphpim,
pim-graphics,
pim-enabled,
Mingyu:PACT,
PICA,
pattnaik.pact16,
cali2020genasm,
syncron,
fernandez2020natsa,
boroumand2021google,
kim.bmc18,
NIM,
pugsley2014ndc,
santos2018processing,
rezaei2020nom,
singh2019napel,
boroumand2021google_arxiv,
oliveira2021pimbench,
}},
non-volatile memories~\cite{ghiasi2022genstore, li.dac16,neurocube,xi2020memory,yavits2021giraf,hamdioui2015memristor})}.\footnote{\gfcri{\gfcrii{Memory} manufacturers \gfcrii{recently} introduced \emph{real} PIM systems that target different application domains (\gfcriii{e.g.,} neural networks~\gfcri{\cite{lee2022isscc,kwon202125,lee2021hardware,ke2021near,niu2022184qps}}, general-purpose computing~\gfcrii{\cite{devaux2019true,gomez2021benchmarking,gomez2021benchmarkingcut}}) and memory technologies (e.g., 3D-stacked DRAM~\gfcri{\cite{kwon202125,lee2021hardware,niu2022184qps}}, 2D DRAM~\gfcrii{\cite{lee2022isscc,ke2021near,gomez2021benchmarking,gomez2021benchmarkingcut}}, non-volatile memories~\cite{Mythic}).}}} We equip the analytical islands with a new \gfcr{PIM-based} analytical engine (\gfcri{\cref{sec:proposal:analytic-engine}}) that includes simple in-order PIM cores added to the logic layer of a 3D-stacked memory~\gfcri{\cite{hmcspec2,kwon202125,lee2016smla}}, software to handle data placement, and runtime task scheduling heuristics. \gfcri{Our new} design enables the execution of transactional and analytical workloads at low latency and high throughput (\emph{third desired HTAP property}).}

\sgii{In our evaluations \am{(\gfcri{\cref{sec:eval})}}, we show the benefits of each component of \titleShort,
and compare its end-to-end performance and energy usage to three
state-of-the-art HTAP systems \sgcri{(modeled after Hyper~\cite{hyper}, AnkerDB~\cite{ankerdb}, and Batch-DB~\cite{batchdb})}.
\titleShort outperforms all three, with higher 
transactional \gfcri{throughput} (2.20$\times$/1.15$\times$/1.94$\times$; mean of 1.70$\times$) and 
analytical \gfcri{throughput} (3.78$\times$/5.04$\times$/2.76$\times$; mean of 3.74$\times$)\gfcr{, while consuming 48\% lower energy than the prior lowest-energy HTAP system}. \gfcr{W}e conclude that \titleShort efficiently provides
high-throughput real-time \gfcri{data} analysis, \gfcrii{while} meeting all three
desired HTAP properties.}

\Copy{CR1/1C}{
\gf{We make the following key contributions in this work:}
\begin{itemize}[leftmargin=1em,nosep]
\item \gf{We comprehensively examine major system- and architecture-level
challenges \rev{that hinder throughput} in HTAP systems.}

\item \gfcri{We propose \titleShort, an HTAP system composed of heterogeneous hardware--software co-designed components (called \emph{islands}) that are specialized for executing transactional and analytical workloads. \sgcri{For each island, we develop software-based optimizations and design dedicated hardware resources (e.g., processing-in-memory-based accelerators for analytical \gfcrii{islands}), both of which} cater to the \sgcri{memory usage and computational properties of their target} workloads.}

\item \gfcri{To achieve all three desired HTAP properties, we co-design \sgcri{algorithm\gfcrii{ic} modifications} and PIM hardware accelerators for update propagation and data consistency\gfcrii{,} aiming to reduce data movement overheads.}

\item \gfcrii{We tailor data placement and task scheduling schemes to the memory characteristics of HTAP workloads, aiming to fully exploit main memory bandwidth and \gfcriii{system} utilization.}

\item \gf{We extensively compare \titleShort against three state-of-the-art HTAP systems\gfcri{. W}e show that \titleShort provides higher transactional and analytical throughput \gfcri{and lower energy} compared to the baseline HTAP systems while meeting all three desired HTAP properties.} 

\item \gfcrii{We open-source \titleShort and the complete source code of our evaluation~\cite{polynesia.github}.}

 \end{itemize}
}
\section{HTAP Background}
\label{sec:bkgd:requirements}

\gfcr{To enable real-time \gfcri{data} analysis, \gfcri{where data analysis needs to be performed using the most recent version of data,} a DBMS needs to be capable of efficiently executing analytics on fresh (i.e., the most recent) version of data that is ingested by transactional workloads, which is a challenging task.} {Several} works from industry (e.g., \cite{sap-hana, oracle-dual-format, sql-htap,sap-soe, real-time-analysis-sql}) 
 and academia (e.g., \cite{hyper,peloton,hyrise,h2tap,l-store,batchdb,scyper, janus,janus-graph,sap-parallel-replication}
  attempt to address \sgmod{this issue and propose}{issues with data freshness by proposing} various techniques to 
support both transactional and analytical workloads in a \emph{single} database system. 
This combined approach is known as \emph{hybrid transactional and analytical processing} (HTAP).
\label{sec:bkgd:htap-requirements}
To enable real-time analysis, an HTAP system should exhibit \gfcr{three} key properties~\cite{batchdb}\gfcrii{.}

\paratitle{\gfcri{Property 1:} Workload-Specific Optimizations}
The \gfcr{HTAP} system should provide \gfcr{transactional and analytical workloads} with optimizations specific to
\gfcr{each of them}. \gfcr{Transactional and analytical workloads} require different algorithms and data structures,
based on the workload's memory access patterns, to achieve high throughput and performance\gfcr{. This leads to different and conflicting optimization techniques (e.g., data layout, hardware design) that can be applied to transactional and analytical workloads}. 
 
\paratitle{\gfcri{Property 2:} Data Freshness and Data Consistency} The \gfcr{HTAP} system should 
provide the analytics workload
 with the \emph{most recent version} of data, even when transactions keep updating the
 data at a high rate. Also, the system needs to guarantee data consistency
 across the entire system, such that analytic\gfcri{al} queries observe a consistent view of data,
regardless of the freshness of the data.

\paratitle{\gfcri{Property 3:} Performance Isolation} The \gfcr{HTAP} system should ensure that the latency
 and throughput of either \gfcri{the} \gfcr{transactional or analytical} workload is not impacted by running them concurrently
 within the same system.

\vspace{3pt}
Meeting all \gfcr{three desired HTAP properties simultaneously} is very challenging~\gfcri{\cite{batchdb,h2tap}}, as \gfcr{transactional and analytical} workloads have different underlying algorithms and access patterns, and optimizing for one property can often require a trade-off
in another property. 

\section{Motivation}
\label{sec:bkgd:motiv}

There are two major types of HTAP systems:
(1)~single-instance design systems and
(2)~multiple-instance design systems.
In this section, we study both types, and analyze why neither type 
can meet all \gfcri{three} desired properties of an HTAP system (\gfcri{see \gfcri{\cref{sec:bkgd:htap-requirements}}).}\gfdel{ as we
describe in Section~\ref{sec:bkgd:htap-requirements}).} \gfcri{To illustrate the key challenges faced by the two types of HTAP systems, we assume a relational DBMS (RDBMS), where data is stored in \sgcri{two-dimensional} tables, with tuples (rows in the table) representing a set of data related to different attributes (columns in the table)~\cite{codd1990relational}.}

\subsection{Single-Instance Design}
\label{sec:bkgd:single}

\gfrev{Single-instance HTAP systems\gfrev{~\cite{hyper, peloton, l-store, hyrise, h2tap, sap-hana}}}\gfdel{One way to design an HTAP system is to} maintain a single instance of the data 
\gfcri{that} both analytics and transactions work on, ensuring that
analytical queries access the most recent version of data. While single-instance design enables high data freshness, it suffers from three major challenges:

\paratitle{(1)~High Cost of Consistency and Synchronization}
\gfcr{Single-instance-based HTAP systems 
need to ensure that the data is consistent and synchronized, since analytical and transactional workloads 
work on the same instance of data concurrently. One approach to consistency is to let both transactions and analytics work on the same copy of data, and use locking
protocols~\cite{locking} to maintain consistency across the system. However,
locking protocols lead to throughput bottlenecks for both transactional and analytical workloads~\cite{virtual-snapshot-nosql,hyper2}.
} 
\sgmod{Since analytical and transactional workloads 
work on the same instance of data concurrently, single-instance-based systems 
need to ensure that the data is consistent and synchronized. One approach to consistency is to let both
transactions and analytics work on the same copy of data, and use locking
protocols~\cite{locking} to maintain consistency across the system. However,
locking has two major drawbacks. First, it significantly reduces the update throughput of transactions~\cite{virtual-snapshot-nosql,hyper2}
and degrades data freshness, as it blocks transactions from updating objects that
 are being read by long-running analytics queries. Second, it can frequently
 blocks analytics, as with the high update rate of transactions,
 a transactional workload can often lock out the analytics workload,
leading to a significant drop in throughput. To avoid these drawbacks,}{To avoid the throughput bottlenecks
incurred by locking protocols~\cite{locking},}
 single-instance HTAP systems resort to either 
 snapshotting~\cite{hyper,h2tap,scyper,ankerdb} or multi-version concurrency
 control (MVCC)~\cite{hyper2, peloton}. Unfortunately, \sgmod{we find that snapshotting and MVCC}{both solutions} have significant drawbacks\gfdel{ of their own}.

\emph{Snapshotting:} Several HTAP systems (e.g., \cite{hyper, scyper, h2tap}) 
use a variation of multiversion synchronization, called snapshotting, to
 provide consistency via snapshot isolation~\cite{isolation-level,serializable-isolation}.
 Snapshot \gfcr{i}solation guarantees that all reads in a
 transaction see a consistent snapshot of the database state, which is
 the last committed state before the transaction started. \gfdel{These systems explicitly
 create snapshots from the most recent version of operational data, and let the analytics
 run on the snapshot while transactions continue updating the data. }

\begin{figure}[h]
    \centering
        \centering
        \includegraphics[width=\linewidth]{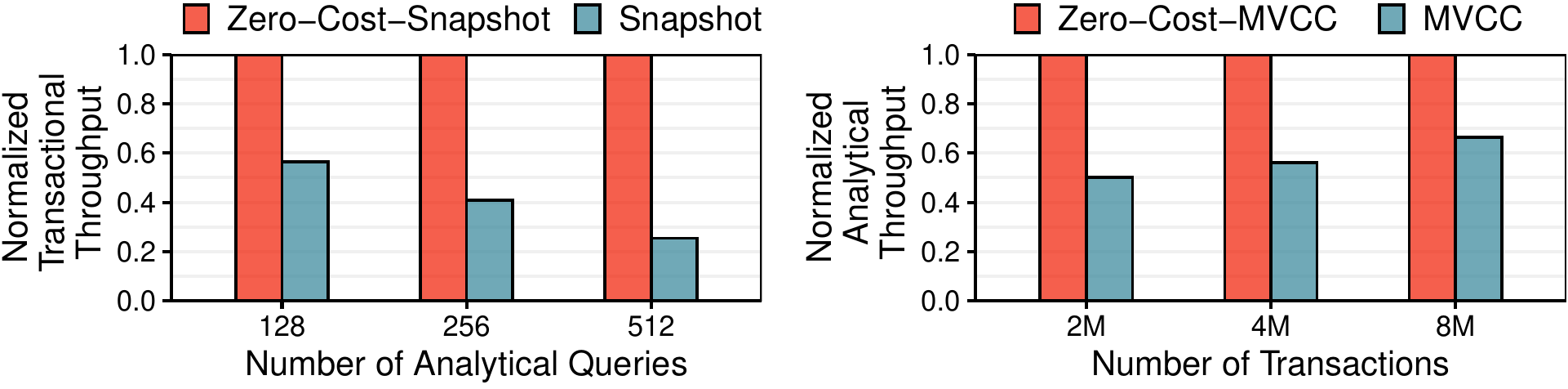}%
    \caption{Effect of \gfcrii{snapshotting on transactional throughput (\gfcrii{left}) and} MVCC on analytical throughput (\gfcrii{right}).}
    \label{fig:motivation-consistency}
\end{figure}

We analyze the effect of state-of-the-art snapshotting~\cite{software-snapshotting,h2tap} 
 on \gfcri{the} throughput \gfcri{of} an HTAP system with two transactional and two analytical
 threads (each run\gfcri{ning} on a separate CPU; see \gfcri{\cref{sec:methodology}} for our evaluation methodology).  \gfcri{Fig.}~\ref{fig:motivation-consistency} (\gfcrii{left})
shows the transaction\gfcriii{al} throughput with snapshotting, normalized to a
zero-cost snapshot mechanism (i.e.,  a hypothetical ideal baseline  where snapshotting operations incur zero delay during execution),  for \gfcri{three} analytical quer\gfcri{y counts}. 
We make two observations.
First, at 128~analytical queries, snapshotting reduces \gfcriii{transactional} throughput by 43.4\%.
Second, the throughput \gfcri{loss increases} as more analytical queries are being performed, with a \gfcri{loss} of
74.6\% for 512~analytical queries.
We find that the majority of this throughput \gfcri{loss} occurs because \texttt{memcpy} is
used to create each snapshot, which introduces significant interference among the
workloads and generates a large amount of data movement between the CPU and main
memory~\gfcri{\cite{seshadri2013rowclone,chang.hpca16}}.
The resulting high contention for shared hardware resources\gfdel{  (e.g., off-chip channel, 
memory system)} directly hurts the throughput.

\emph{MVCC:} \gfcri{In MVCC, instead of replacing the old data in a tuple when updates happen (as in the snapshotting approach), the system creates a new timestamped version of the data that is chained together with old versions of the data, forming a pointer-based \emph{version chain}. During execution, instead of reading a separate snapshot, an analytical query can simply use a timestamp to read the most up-to-date version of the data \sgcri{at the time the query starts. Concurrently}, a transactional query can insert transactional updates \sgcri{at} the end of the version chain \sgcri{with more recent timestamps, without affecting the consistency of the analytical query}. As a result, updates never block reads, which is the main reason why many transactional DBMSs have adopted MVCC (e.g., \cite{sap-hana,oracle-dual-format,sql-htap}).}

However, MVCC is not a good fit for mixed analytical and transactional workloads in HTAP. We study the effect of MVCC on system throughput, using
the same hardware configuration that we used for snapshotting. \gfcri{Fig.}~\ref{fig:motivation-consistency} (\gfcrii{right}) shows 
the analytical throughput of MVCC, normalized to a zero-cost version of MVCC (i.e.,  a hypothetical ideal baseline  where MVCC operations incur zero delay during execution),
for \gfcri{three} transactional query count\gfcri{s}. We observe that the analytical throughput significantly decreases \unsure{Why does throughput loss reduce as number of Transactions increases?}(by 42.4\% \gfcri{on average across the three transactional query counts}) compared to zero-cost MVCC. \gfcri{The root cause is the long version chain caused by frequent updates from the transactional workload. Each version chain is organized as a linked list, which grows with the number of updates from transactional queries. When accessing data in a tuple, the analytical query needs to traverse a lengthy version chain, checking the timestamp in each element of the linked list to locate the most recent version of the data. As analytic\gfcri{al} queries touch large
 volumes of data, this generates a large number of random memory accesses, leading to the \gfcrii{large} throughput \gfcri{loss}.}
 
\paratitle{(2)~Limited Workload-Specific Optimization\gfcri{s}} 
A single-instance design severely limits workload-specific optimizations, as the instance cannot have different optimizations for each workload. For example, relational transactional 
 engines use a row-wise or N-ary storage model (NSM) for data layout~\cite{kallman2008h}\gfdel{,
 as it provides low latency and high throughput for update-intensive queries~\cite{kallman2008h}.
 R}\gfrev{, while r}elational analytics engines employ a column-wise or decomposition storage model (DSM)~\gfcri{\cite{copeland1985decomposition,c-store}.} It is inherently impossible for a single-instance-based system to \gfrev{efficiently} implement both 
\gfcri{data layouts} simultaneously, and many such systems simply choose one of the 
layouts~\cite{hyper, l-store, h2tap}.

\paratitle{(3)~Limited Performance Isolation}
We evaluate \gfdel{the effect of} performance
interference using the same system configuration that we used for snapshotting and MVCC. Each transactional thread executes
2M queries, and each analytical thread runs 1024 analytical queries. 
We assume that there is no cost for consistency and synchronization.
Compared to running transactional queries in isolation, the transactional throughput
drops by 31.3\% when the queries run alongside analytics. This is because analytics are very data-intensive and generate a large
 amount of data movement, which leads to significant 
 contention for shared resources (e.g., \gfcriv{the} memory system~\gfcriv{\cite{subramanian2013mise,das2013application,mutlu2007stall,mutlu2008parallelism,MPKI-ref2,ebrahimi2010fairness,mutlu2013memory,mutlu2015research,mutlu2015main,subramanian2015application,subramanian2015providing, atlas,TCM,mutlu2007memory,usui2016dash,subramanian2016bliss,lee2015decoupled,ausavarungnirun2012staged,grot2011kilo,nesbit2006fair,nesbit2008multicore}}). Note that the problem worsens with realistic consistency mechanisms, as they also generate a large amount of data movement. 

\subsection{Multiple-Instance Design}
\label{sec:bkgd:multiple}

\gfcri{A second major} approach to design\gfcrii{ing} an HTAP system is to maintain multiple instances of the data using
 replication techniques~\gfcri{\cite{mvcc,software-snapshotting}}, and \gfcrii{specialize} each instance to a specific
 workload (e.g., \cite{batchdb,oracle-dual-format,sql-htap,sap-soe,scyper,janus,htap-survey}). Unfortunately, multiple-instance \sgdel{design }systems
suffer from \sgmod{a number of}{several} challenges\gfcri{.}

\paratitle{Data Freshness} One of the major challenges in \gfcri{the} multiple-instance \gfcri{design} 
is keep\gfcri{ing} analytical replicas up-to-date even when the transaction update rate is high, 
without compromising performance isolation~\cite{htap,batchdb}. To maintain data freshness, the
system needs to \gfcri{propagate transactional updates to analytical replicas (referred as \emph{update propagation}), which requires}
(1)~gather\gfcri{ing} updates from transactions and ship\gfcri{ping} them to
analytical replicas \gfcri{(\emph{update gathering and shipping})}, and 
(2)~perform\gfcri{ing} the necessary format conversion and \gfcri{applying} the
updates \gfcri{(\emph{update \gfcrii{application}}).} \sgcri{As we discuss below, resource contention and data movement costs become significant performance limiters for multiple-instance HTAP systems.}

\emph{\gfcri{Update Gathering and Shipping}:} 
Given the high update rate of transactions, the frequency of the gathering and shipping process has a direct effect on data freshness. During \sgcri{this} process, the system needs to (1)~gather updates from different transactional threads, (2)~scan them to identify the target \gfcri{memory} location corresponding to each update, and (3)~transfer each update to the corresponding \gfcri{memory} location. 

\gfcri{\emph{Update Application}: The update application process can be challenging due to the need to transform updates from one workload-specific format to another. In RDBMSs, analytical engines use DSM representation to store data~\cite{c-store}  and can compress tuples using order-preserving dictionary-based compression (e.g., dictionary encoding~\cite{scaling-up-c-store-numa,dict-compression,compression-c-store}) \gfrev{t}o minimize the amount of data that needs to be accessed. In contrast, a single tuple update, stored in the NSM layout by the transactional workload, requires multiple random \gfcri{memory} accesses to apply the update in the DSM layout. Compression further complicates this, as columns may need to be decompressed, updated, and recompressed. For compression algorithms that use sorted tuples, such as dictionary encoding, the updates can also lead to expensive shifting of tuples. These operations generate a large amount of data movement and \gfcri{consume} many CPU cycles. The challenges are significant enough that some prior works give up on workload-specific optimization to maintain reasonable system performance~\cite{batchdb}.}

\begin{figure}[h]
    \centering
        \centering
        \includegraphics[width=\linewidth]{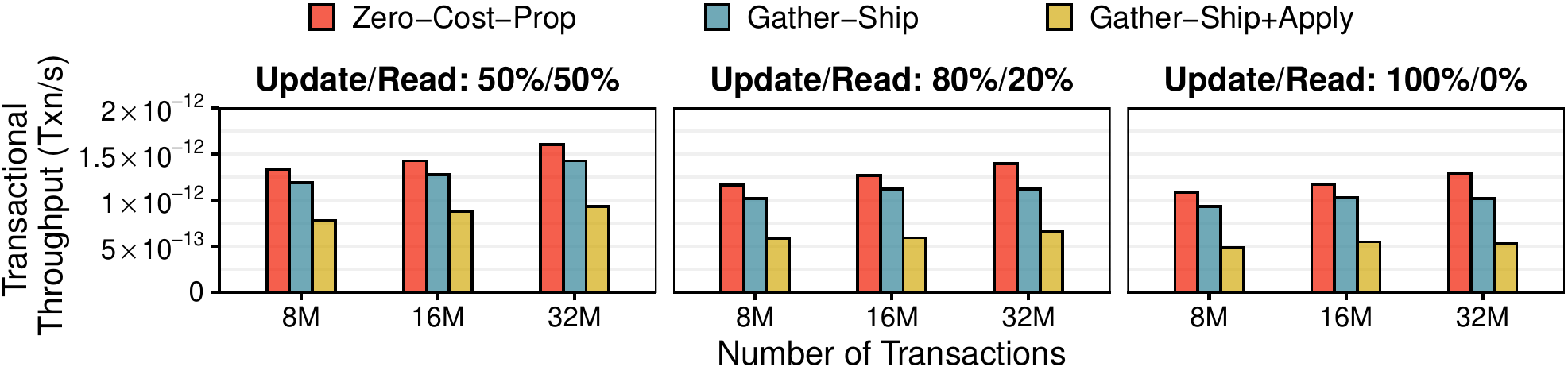}%
    \caption{Transactional throughput across different number\gfcri{s} of transactions and different \gfcri{update} intensities.}
    \label{fig:motivation-update-propagation}
\end{figure}

\gfcri{We study the effect of \gfcri{update propagation} \gfcrii{(i.e, update gathering and shipping, and application)} on \gfcri{the} transactional throughput \gfcri{of} a multiple-instance HTAP system  (see \gfcri{\cref{sec:methodology}}).  \gfcri{Fig.}~\ref{fig:motivation-update-propagation} shows the transactional throughput for three configurations: 
(1)~a baseline system with \sgcri{zero cost} for \gfcri{update propagation} \gfcri{(\emph{Zero-Cost-Prop})},
(2)~a system that performs \emph{only} update gathering and shipping \gfcri{(\emph{Gather-Ship})}, and 
(3)~a system that performs \gfcri{\sgcri{update gathering, shipping, \emph{and} application} \gfcri{(\emph{Gather-Ship+Apply})}}. Our system has two transactional \sgcri{threads} and two analytical threads (each running on a CPU core) in all three configuration\gfcrii{s}. We make two observations. First, we observe a loss in transactional throughput due to the update gathering and shipping process, which increases as a factor of the update intensity of the transactional query. The transactional throughput of the \emph{Gather-Ship} configuration \gfcrii{is} 11\% \gfcrii{lower than} that of the \emph{Zero-Cost-Prop} configuration for a 50\% update intensity, on average, across different transaction counts. When the transactional queries are more update-intensive (e.g., 80\% to 100\% updates), the overhead becomes significantly higher, with a throughput loss of 19.9\% and 21.2\% for 80\% and 100\% update intensities, respectively. Second, we observe that the update application process leads to an additional loss in transactional throughput. The transactional throughput of the \emph{Gather-Ship+Apply} configuration reduces by 41\%, on average, across different transaction counts, compared to that of the \gfcri{\emph{Zero-Cost-Prop}} configuration for a 50\% update intensity. As the \gfcri{update} intensity increases (from 50\% to 80\%), the loss in transactional throughput further increases (with a 59.0\% \gfcrii{loss} at 80\% update intensity).}

\begin{figure}[h]
    \centering
        \centering
        \includegraphics[width=\linewidth]{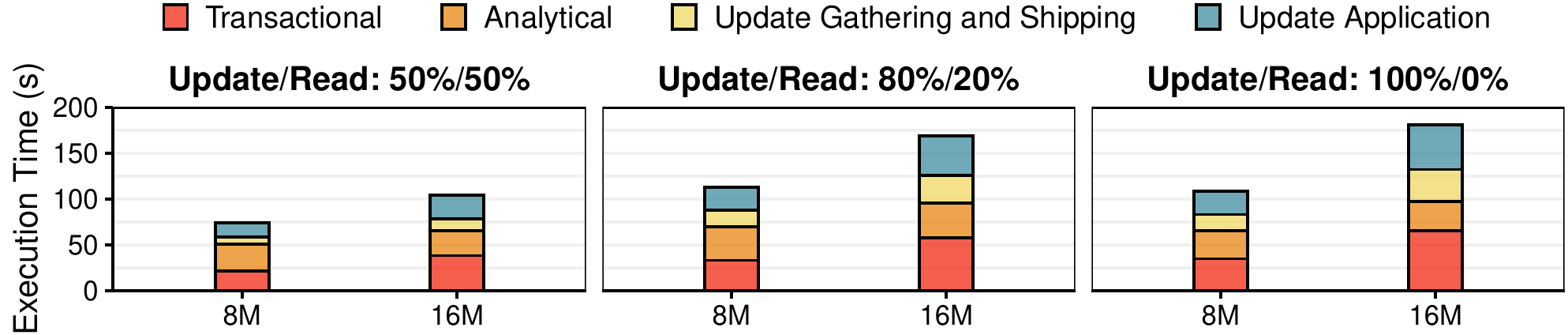}%
    \caption{Execution time breakdown across different number\gfcri{s} of transactions \gfcri{update} intensities.}
    \label{fig:motivation-update-propagation-latency}
\end{figure}

\gfcri{We further analyze the impact of update propagation on the execution time of our HTAP system. \gfcri{Fig.}~\ref{fig:motivation-update-propagation-latency} 
shows the breakdown of execution time during \gfcri{the} update propagation process for different numbers of transactions and different update intensities. We make two observations. 
First, update gathering and shipping 
accounts for 15.4\% of the total execution time, on average, which stems from the large amount of data movement generated by the update gathering and shipping process. Second, the update application process accounts for 23.8\% of the execution time, of which 62.6\% is spent on (de)compressing columns. Our analysis shows that similar to update gathering and shipping, the update application process also suffers data movement overheads since 30.8\% of the total last-level cache misses, on average, are generated during the update application process. We conclude that update propagation accounts for a significant portion of the execution time in our HTAP system (39\%, on average), resulting in the loss in transactional throughput \sgcri{that} we observe in \gfcri{Fig.}~\ref{fig:motivation-update-propagation}.}

\paratitle{Other Major Challenges} 
\gfcri{W}e find that maintaining data consistency for multiple instances without compromising
performance isolation is very challenging. Updates from transactions are frequently shipped
and applied to analytic\gfcri{al} replicas while analytical queries run. 
As a result, multiple-instance systems suffer from the same consistency drawbacks
that we observe for single-instance systems in \gfcri{\cref{sec:bkgd:single}}. Another major challenge we find is the limited performance isolation. While separate
 instances provide partial performance isolation, as \sgcri{transaction\gfcrii{al} queries and analytical queries} do not compete for
 the same copy of data, they still share \gfcri{and contend for} underlying hardware resources such as CPU cores and 
 the memory system~\gfcriv{\cite{subramanian2013mise,das2013application,mutlu2007stall,mutlu2008parallelism,MPKI-ref2,ebrahimi2010fairness,mutlu2013memory,mutlu2015research,mutlu2015main,subramanian2015application,subramanian2015providing, atlas,TCM,mutlu2007memory,usui2016dash,subramanian2016bliss,lee2015decoupled,ausavarungnirun2012staged,grot2011kilo,nesbit2006fair,nesbit2008multicore}}. \gfdel{As we discuss in Section~\ref{sec:bkgd:single},
 analytics workloads, as well as data freshness and consistency mechanisms, generate a large amount of data movement and take many cycles. As a result, multiple-instance designs also suffer from limited performance isolation. }

\vspace{3pt}
We conclude that neither single- nor multiple-instance HTAP systems meet \gfcri{the three} desired HTAP properties \gfcri{(\gfcri{\cref{sec:bkgd:htap-requirements}})}. We, therefore, need a new system that can avoid resource contention and alleviate the data movement costs incurred in HTAP systems.

\section{\texorpdfstring{\MakeUppercase\titleShort}{\titleShort}}
\label{sec:proposal}

\sgmod{Our goal in this work is to design an HTAP system that can meet 
all of the desired HTAP properties, by avoiding the challenges that we
identify for state-of-the-art HTAP systems in \gfcri{\cref{sec:bkgd:motiv}}.
To this end, we propose \titleLong (\titleShort).
\titleShort}{We propose \titleShort, which} divides the HTAP system into multiple \emph{islands}. \gfcri{An island is a hardware--software co-designed component specialized for \gfcrii{specific} types of queries.} Each island includes (1)~a replica of data whose layout is optimized for a specific workload, 
(2)~an optimized execution engine, and (3)~a set of hardware \sgmod{resources (e.g., computation units, memory).}{resources.}
 \titleShort has two types of islands: 
(1)~a \emph{transactional island}, and 
(2)~an \emph{analytical island}. \rev{To avoid the data movement and interference
challenges that other multiple-instance HTAP systems face (see
\gfcri{\cref{sec:bkgd:motiv}}),} we propose to equip each analytical island with
(1)~\emph{\gfcri{processing-}in-memory \gfcri{(PIM)} hardware}; and 
(2)~co-designed algorithms and hardware \gfcri{to execute analytical workloads} \gfcriii{as well as to} \gfcrii{perform}
\emph{update propagation} and \gfcriii{to guarantee data} \emph{consistency}.

\begin{figure}[ht]
    \centering
        \centering
        \includegraphics[width=\linewidth]{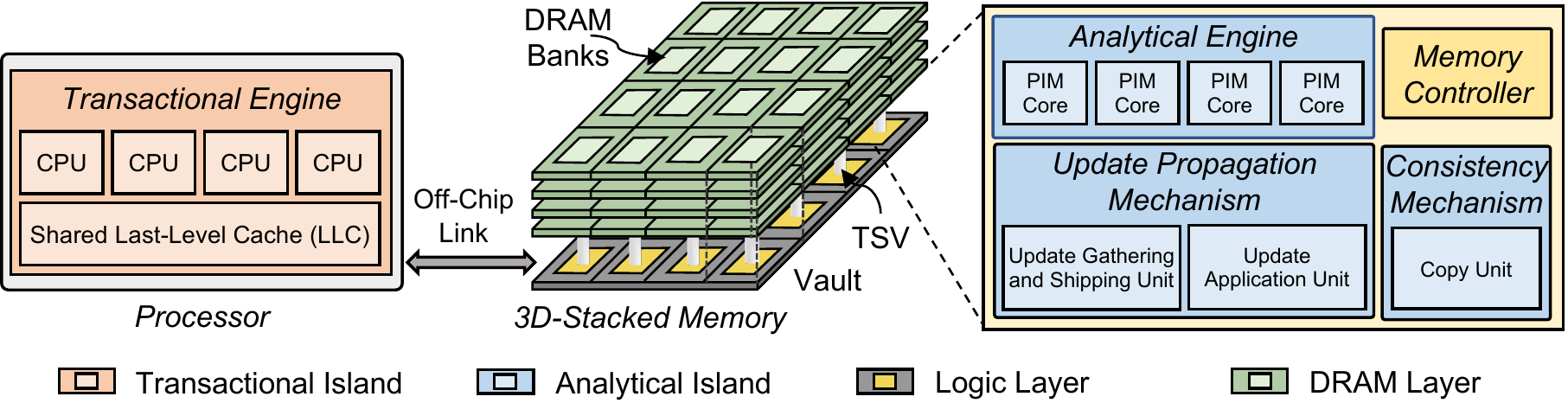}%
    \caption{High-level organization of \titleShort hardware.}
    \label{fig:high-level-hw}
\end{figure}

\titleShort is a framework that can be applied to many different combinations
of transactional and analytical workloads.
 In this work, we focus on designing an instance of \titleShort 
 that supports relational transactional and analytical workloads.\footnote{Note that our proposed techniques
 can be applied to other types of analytical workloads (e.g., graphs, machine learning) as well.}
\gfcri{Fig.}~\ref{fig:high-level-hw} shows the hardware for our chosen implementation,
which includes one transactional island and one analytical island, and is equipped with
a 3D-stacked memory similar to the Hybrid Memory Cube (HMC)~\cite{hmcspec2},
where multiple vertically-stacked DRAM layers are connected with a 
\emph{logic layer} using thousands of \emph{through-silicon vias} (TSVs). 
An HMC chip is split up into multiple \emph{vaults}, where each vault corresponds to a vertical slice of the memory and logic layer. 

\gfcri{\titleShort's} transactional island uses an execution engine similar to conventional transactional engines~\cite{dbx1000,kallman2008h} to execute a relational  transactional workload. The transactional island is equipped with conventional multicore CPUs and multi-level caches, as transactional queries have
 short execution times, are latency-sensitive, and have cache-friendly access patterns~\gfcriv{\cite{conda,LazyPIM,amiraliphd}}. \gfcri{\titleShort's analytical island uses specialized PIM hardware.} Inside each vault's portion of the logic layer in memory, we add hardware for the analytical island, including the update propagation mechanism (consisting of the \emph{update \gfcri{gathering and} shipping} \sgcri{unit} and \emph{update application} \sgcri{unit}), the consistency mechanism (\emph{copy \sgcri{unit}}), and the \emph{analytical execution engine} (simple programmable in-order PIM cores).\footnote{\gfcr{T}he hardware components of Polynesia are a combination of general-purpose processors (GPPs) and fixed-function (ASIC) components. The GPP is responsible for executing transactional queries (at the host processor side) \gfcr{and} analytical queries (from within the 3D-stacked DRAM device). The ASIC components are responsible for executing the update propagation and consistency mechanisms we propose. Our current ASIC designs are \emph{not} reconfigurable to ease implementation, but could be extended to be reconfigurable.}%

\sg{\gfcri{In the next sections, we discuss the detailed design of \titleShort's main components, which includes the update propagation mechanism (\cref{sec:proposal:update-propagation}), the consistency mechanism (\cref{sec:proposal:consistency}), and the analytical execution engine (\cref{sec:proposal:analytic-engine}). We discuss \emph{both} algorithmic optimizations and novel hardware \sgcri{for} each component.}}
\gfdel{\sgiv{To address potential capacity issues and accommodate larger data, 
\titleShort can extend across multiple memory stacks.  We evaluate
\titleShort with multiple stacks in Section~\ref{sec:eval:multiple}.}}

\section{Update Propagation Mechanism}
\label{sec:proposal:update-propagation}

We design \sgii{a new two-part} update propagation mechanism\gfdel{ to overcome the high costs
\sgii{of analytical replica updates in state-of-the-art HTAP systems}}. The \emph{update \gfcri{gathering and} shipping unit} gathers updates from the transactional island,
finds the target location in the analytical island, and frequently 
pushes these updates to the analytical island.
The \emph{update application unit} receives these updates,
converts the updates from the transactional to the analytical replica data format, and applies the update to the
analytical replica. \gfcrii{Our update propagation mechanism leverages novel algorithms and hardware accelerators tailored to reduce data movement overheads while maintaining data freshness between transactional and analytical islands.}

\subsection{Update \gfcri{Gathering and} Shipping}
\label{sec:proposal:update-propagation:update-shipping}

\paratitle{Algorithm} 
\rev{Our update \gfcri{gathering and} shipping mechanism includes three major stages. For each thread in} the transactional engine, \titleShort stores an ordered \emph{update log}
for the queries performed by the thread. Each update log entry contains four fields: 
(1)~a commit ID (a timestamp used to track the total order of all
updates across threads), (2)~the type
 of the update (insert, delete, modify), (3)~the updated data, and (4)~a record key (e.g., pair of
 row-ID and column-ID) that links this particular update to a column in
the analytic\gfcri{al} replica. The update \gfcri{gathering and} shipping process is triggered when \gfcriii{the} total number of pending updates reaches the final log
 capacity, which we set to 1024 entries (see \gfcri{\cref{sec:proposal:update-propagation:update-application}}).
\sgcri{Stage~1 scans} the per-thread update logs,
and \sgcri{merges} them into a single \emph{final log}, where all updates are sorted by the commit ID.

 \sgcri{Stage~2 finds} the \gfcri{memory} location of the corresponding column (in the analytical replica) 
 associated with each update log entry. \rev{We observe that this stage is one of the major bottlenecks of update \gfcri{gathering and} shipping, because the
 fields in each tuple in the transactional island are distributed across different columns
 in the analytical island. Since the column size is typically very large, 
 finding the \gfcri{memory} location of each update is a very time-consuming process. To overcome this,} we maintain a hash index \sgcri{of the stored data} on the \texttt{(column,row)} key, and
 use that to find the corresponding column\gfcrii{'s} \gfcri{location} for each update
 in the final log. \rev{We use the \texttt{modulo} operation as the hash function. \gfcri{Our hash table uses bucket hashing with separate chaining to handle collisions, and hash buckets \sgcri{containing a linked list of keys} are stored in main memory}. We size our hash table based on
 the column partition size.\footnote{Similar to conventional analytical DBMSs, we can
 use soft partitioning~\cite{batchdb, morsel, scaling-up-c-store-numa} to address scalability
 issues when the column size increases beyond a certain point. Thus, the hash table
 size does \emph{not} scale with column size.}}
 \gfcri{We place the updates from the final log for each column in a \emph{column buffer}, based on the output of the hash unit.  At the end of this stage, there are multiple column buffers, each corresponding to a column in the analytical replica, which are ready to be shipped (i.e., written) to the analytical island.} \sgcri{Stage~3 ships} \gfcri{all column} buffers to each column in the analytical replica. 
 
\paratitle{Hardware}
\rev{We find that despite our best efforts to minimize overheads, our algorithm
has three major bottlenecks that
 keep it from meeting data freshness and performance isolation requirements:} 
(1)~the scan and merge \rev{operation} in \sgcri{Stage~1},
(2)~hash index lookups in \sgcri{Stage~2}, and
(3)~transferring the column buffer contents to the analytical islands in \sgcri{Stage~3}.
\rev{These primitives generate a large amount of data movement and account for 87.2\% of
 our algorithm's execution time.} To avoid these bottlenecks, we \sgii{design a new hardware accelerator, 
called the \emph{update \gfcri{gathering and} shipping unit}, \rev{that speeds up the key primitives of the update \gfcri{gathering and} shipping algorithm.}
We add \rev{this accelerator} to each of
\titleShort's in-memory analytical islands}.

\begin{figure}[ht]
    \centering
    \includegraphics[width=\linewidth]{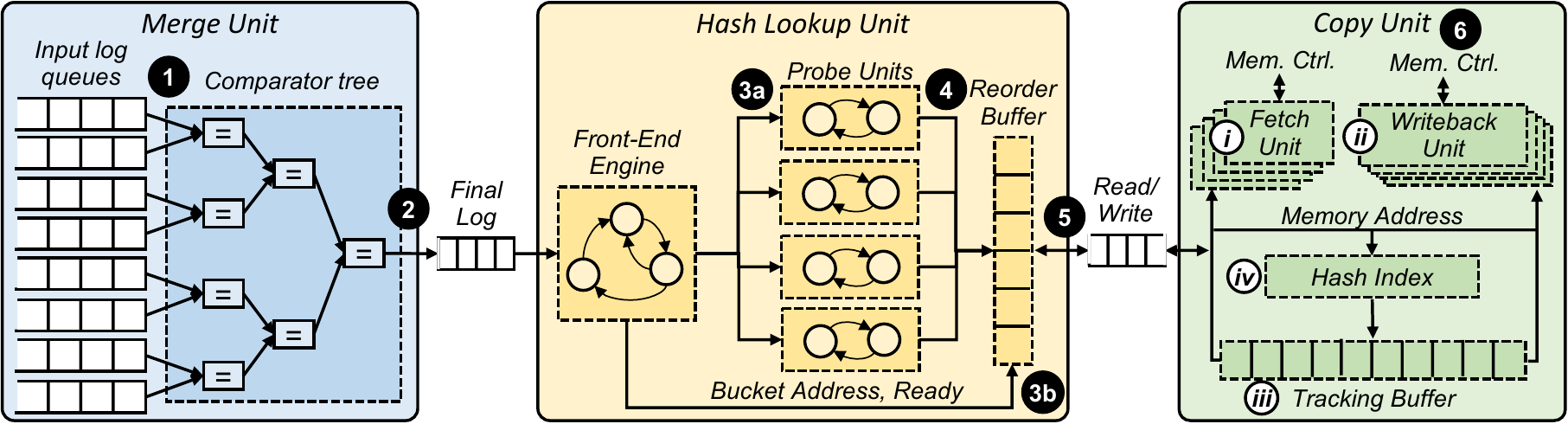}%
    \caption{\rev{Update \gfcri{gathering and} shipping unit architecture.}}
    \label{fig:update-shipping-hw}
\end{figure}

\rev{\gfcri{Fig.}~\ref{fig:update-shipping-hw} shows the high-level architecture of our in-memory 
update \gfcri{gathering and} shipping unit.}
The update \gfcri{gathering and} shipping unit consists of three building 
blocks: 
(1)~a merge unit, \gfcri{which merges the per-thread sorted update logs into the single final log (\sgcri{Stage~1} of our algorithm);} 
(2)~a hash lookup unit, \gfcri{which decouples hash lookup operations into two steps, \gfcrii{i.e.,} bucket address generation and bucket access and traversal (\sgcri{Stage~2} of our algorithm)}; and 
(3)~a copy unit, \gfcri{which issues \gfcrii{concurrent read/write} memory requests to main memory (\gfcrii{accelerating data copy operations required during} \gfcrii{hash table indexing} in \sgcri{Stage~2} \gfcrii{and column buffer shipping in Stage~3} of our algorithm)}. 

\gfcri{The \emph{merge unit} consists of 8 FIFO input queues, where each input queue corresponds to a sorted update log. Each input queue can hold up to 128 updates, which are streamed from DRAM. The \emph{hash lookup unit} consists of a front-end engine (a finite-state machine, \sgcri{or FSM,} responsible for bucket memory address generation), four probe units (FSMs responsible for bucket access and traversal), and a small reorder buffer (to track in-flight hash lookups issued to main memory). The hash lookup unit (1)~decouples key hashing and bucket address generation from the actual bucket access/traversal to allow for concurrent hashing operations; and (2)~\gfcrii{guarantees that updates remain in the same order as executed by the transactional engine, by using} a small reorder buffer to maintain \gfcrii{sequential} commit order for completed hash lookups that are sent to the copy engine. The \emph{copy unit} consists of a \gfcrii{read/write} tracking buffer and multiple fetches and write units (we describe our copy unit in detail in \cref{sec:proposal:consistency}). 
}

\gfcri{The update gathering and shipping unit works in five steps. First, the merge unit finds the oldest entry among the input log queue heads using a 3-level comparator tree \gfcrii{(\incircle{1} in Fig.~\ref{fig:update-shipping-hw})} and adds it to the tail of the final log, which consists of a ninth FIFO queue. Second, the final log \gfcrii{(\incircle{2})} is sent to the hash lookup unit to determine the target column address for each update. Third, the front-end engine in the hash lookup unit performs two operations in parallel: it (i) fetches one hash key from the final update log, computes the hash function, and sends the hash key address to probe units \gfcriii{(\incircleddd{3a})}; and (ii) allocates an entry (with the bucket address and a ready bit) in the reorder buffer, for each hash lookup  \gfcriii{(\incircleddd{3b})}. Fourth, a free probe unit \gfcrii{(\incircle{4})} takes the bucket address received from the front-end engine, and issues read requests to the copy unit in order to traverse the \sgcri{bucket's linked list of keys} \gfcrii{(\incircle{5})}. Once the probe unit reaches the end of the bucket linked list, it will hold the corresponding column for an entry in the final update log. Fifth, the probe unit issues write operations \gfcrii{(\incircle{6})}, using the column memory address retrieved from the hash to the copy unit in order to transfer the update in the final log to the target column buffer. This completes the update gathering and shipping process for one entry in the final log.}

\subsection{Update Application}
\label{sec:proposal:update-propagation:update-application}

Similar to other relational analytical DBMSs, our
analytical engine uses the DSM data layout \gfcrii{(see~\cref{sec:bkgd:single})} and 
\emph{dictionary encoding}~\cite{c-store,update-c-store,sap-hana,compression-gpu,scaling-up-c-store-numa,dict-compression,compression-c-store}. With dictionary encoding, each column in a table is transformed into a compressed column using
encoded fixed-length integer values, and a dictionary stores a sorted mapping of
\gfcrii{unencoded} values to encoded values. As we discuss in \gfcri{\cref{sec:bkgd:multiple}}, the layout conversion (our transactional island uses NSM)
and column compression make \gfcri{the} update application process challenging. \sgii{We design a new update application mechanism for \titleShort 
that uses hardware--software co-design to address these challenges.}

\paratitle{Algorithm} 
\sgii{We first discuss an initial algorithm that we develop for update application.} We assume \gfcri{that} each column has \emph{n} entries, and that we have \emph{m} \sgdel{number of }updates. \gfcri{The algorithm has four steps.}  \gfcrii{In \emph{Step 1,}} the algorithm decompresses the encoded column by scanning the column and
looking up in the dictionary to decode each item. This requires \emph{n random accesses} to the dictionary. \gfcrii{In \emph{Step 2,}} the algorithm applies updates to the
decoded column one by one. \gfcrii{In \emph{Step 3,}} it constructs a new dictionary, by sorting the updated column 
and calculating the number of fixed-length integer bits required to encode the
sorted column. Dictionary construction is computationally expensive ($\mathcal{O}((n+m)\log{}(n+m))$\gfcrii{; where \emph{m} is the number of updates \gfcriii{to a column} and \emph{n} is the number of entries in \gfcriii{a} column}) because we need to sort the entire column. \gfcrii{In \emph{Step 4,}} the algorithm compresses the new column
using our \sgmod{newly-constructed}{new} dictionary. While entry decoding happens in constant time, encoding requires a
logarithmic complexity search through the dictionary (since the dictionary is sorted). 

This \sgii{initial} algorithm is memory intensive (Steps~1, 2, 4)
and computationally expensive (Step~3). Having hardware support is \emph{critical} to enabl\gfcrii{ing}
low-latency update application and performance isolation. 
While \sgmod{processing-in-memory (PIM)}{PIM} may be able to help,
our \sgii{initial} algorithm is not well-suited for PIM for two reasons,
and we \sgii{optimize} the algorithm to address both \gfcrii{reasons}.

\emph{Optimization~1: Two-Stage Dictionary Construction.}
\sgii{We eliminate column sorting from Step~3, as it is
computationally expensive.} Prior work~\cite{pim-sort, q100} shows that to efficiently sort more than 1024 values in hardware,
we should provide a hardware partitioner to split the values into
multiple chunks, and then use a sorter unit to sort chunks one at a time.
This requires an area of \SI{1.13}{\milli\meter\squared}~\cite{pim-sort, q100}. \sgii{Unfortunately, since tables can have millions of entries~\cite{update-c-store},
we would need \gfcriii{tens to hundreds of} sorter units to construct a new dictionary,
easily exceeding the total area budget of \SI{4.4}{\milli\meter\squared}
per vault \gfcri{in our baseline 3D-stacked DRAM}~\cite{google-pim,tetris,mondrian}.}

\sgii{To eliminate column sorting, we sort \emph{only} the dictionary, 
leveraging} the fact that
(1)~the existing dictionary is already sorted, and
(2)~the new updates are limited to 1024~values.
\sgmod{As a result, our}{Our} optimized algorithm initially builds a sorted dictionary for
only the updates, which requires a single hardware \sgmod{sorter.
A 1024-value bitonic sorter requires a much smaller area of only
\SI{0.18}{\milli\meter\squared}~\cite{q100}.}{sorter (a 1024-value bitonic sorter 
with an area of only \SI{0.18}{\milli\meter\squared}~\cite{q100}).}
Once the update dictionary is constructed, we now have two sorted dictionaries:
the old dictionary and the update dictionary.
We merge these into a single dictionary using a linear scan ($\mathcal{O}(n+m)$\gfcrii{; where \emph{m} is the number of updates \gfcriii{to a column} and \emph{n} is the number of entries in \gfcriii{a} column}), 
and then calculate the number of bits required to encode the new dictionary.

\emph{Optimization~2: Reducing Random \gfcri{Memory} Accesses.}
To reduce \sgii{the algorithm's memory intensity (which is a result of random \gfcri{memory} lookups),}
we maintain a hash index that links the old encoded value in a column to the new encoded
value. 
This avoids the need to decompress the column and add updates, eliminating
data movement and random \gfcri{memory} accesses for Steps~1 and 2, while reducing the
number of dictionary lookups required \sgii{for Step~4.} The only remaining random \gfcri{memory} accesses are for Step~4, which decrease from $\mathcal{O}((n+m)\log{}(n+m))$
to $\mathcal{O}(n+m)$.

\gfcri{Our \gfcrii{optimized} algorithm has three steps.} \gfcri{First, we} sort the updates to construct the update dictionary. \gfcri{Second, w}e merge the old dictionary and the update dictionary to construct the new dictionary
and hash index. \gfcri{Third}, we use the \gfcri{hash} index and the new dictionary to find the new encoded
value for each entry in the column.

\paratitle{Hardware} 
We \sgii{design a hardware implementation of our optimized algorithm,
called the \emph{update application unit}, and add it to each in-memory
analytical island.} The unit consists of three building blocks: a \emph{sort unit}, a \emph{hash lookup unit}, and
a \emph{scan/merge unit}. 
\sgii{Our sort unit uses} a 1024-value bitonic sorter, \rev{whose basic building block is a network of comparators. These comparators are
 used to form \emph{bitonic sequences}, sequences where the first half of the sequence is monotonically
 increasing and the second half is \gfcri{monotonically} decreasing.
The hash lookup uses a simpler version of the component that we designed for \gfcri{the} update \gfcri{gathering and} shipping \gfcri{unit}.
The simplified version does not use a reorder buffer, as there is no dependency between hash lookups
 for update application. We use the same number of hash units (empirically set to 4), each corresponding to one index structure, to parallelize the compression
 process.} For the merge unit, we use a similar design from our update \gfcri{gathering and} shipping unit. 
\gfdel{Our analysis shows that the total area of our update
application unit is \SI{0.4}{\milli\meter\squared}.}

\section{Consistency Mechanism}
\label{sec:proposal:consistency}

We design a new consistency mechanism for \titleShort in order not to compromise either the throughput of
analytical queries or the \info{Mostly rate.}rate at which updates are applied.  This sets two
\am{requirements} for our mechanism:
(1)~analytical queries must be able to run \gfcri{continuously} without slowdown; and
(2)~the update application process should not be blocked by long-running
analytical queries. 
This means that our mechanism needs a way to allow analytical queries
to run concurrently with updates, without incurring the long\gfcri{-}chain read
overheads of similar mechanisms such as MVCC (see \gfcri{\cref{sec:bkgd:single}}). \gfcrii{Our consistency unit relies on our novel in-memory hardware \emph{copy unit}, which can fully exploit the large internal memory bandwidth of 3D-stacked memories~\gfcriii{\cite{lee2016smla}}.}

\paratitle{Algorithm} 
Our mechanism relies on a combination of snapshotting~\cite{hyper} and versioning~\cite{mvcc} 
to provide snapshot isolation~\cite{serializable-isolation,isolation-level} for analytical queries. 
Our consistency
mechanism is based on two key observations: (1)~updates are applied at a column
granularity, and (2)~snapshotting a column is cost-effective using PIM logic. We assume
that for each column, there is a chain of snapshots where each chain entry corresponds to
a version of \gfcri{the} column. Unlike \gfcri{version} chains in MVCC, each version is associated with a column, not a tuple. 

We adopt a lazy approach (late materialization~\cite{abadi2007materialization}), where \titleShort does not create a snapshot
every time a column is updated. 
Instead, \sgii{on a column update}, \rev{\titleShort marks} the column as dirty, indicating that
the snapshot chain does not contain the most recent version of the column data.
\sgii{When an analytical query arrives\gfcriii{,} \titleShort checks the column \rev{metadata}, and 
creates a new snapshot only if \info{Correct.}\am{(1)~any of the columns are dirty (similar to Hyper~\cite{hyper}), and (2)~no 
\sgiii{current} snapshot exists for the same column (\sgiii{we let multiple queries share a single snapshot}).}
During snapshotting, \titleShort} updates the head of the snapshot chain with the new value, and \rev{marks} the column as clean.
This \sgii{provides two benefits.
First, the analytical query avoids the \gfcri{version} chain traversals and timestamp comparisons performed in MVCC,
as the query only needs to access the head of the \gfcri{version} chain at the time of the snapshot. Second, \rev{\titleShort uses \am{simple yet} efficient garbage collection:} when an analytical query finishes,
snapshots no longer in use by any query are deleted 
(\gfcrii{except for} the head of the \gfcri{snapshot} chain).}

 To maintain high data freshness, our
 consistency mechanism always \gfcrii{allows} transactional updates \gfcrii{to} directly
update the main replica 
\rev{using our two-phase update application algorithm
(\gfcri{\cref{sec:proposal:update-propagation:update-application}}).
In Phase~1, the algorithm constructs a new
dictionary and a new column.
In Phase~2, the algorithm atomically updates the main replica with pointers to the
new column and dictionary.}

\paratitle{Hardware}
\am{\sgiii{Our algorithm's success at} satisfying
 the \sgiii{first requirement for a consistency mechanism (i.e., no slowdown for
analytical queries) relies heavily on its ability to perform fast memory copies
to minimize the snapshotting latency.}
\sgiii{Therefore, we} \sgii{add a custom \gfcrii{\emph{copy unit}} to
each of \titleShort's in-memory analytical \sgiii{islands.}} \sgiii{We have two design
 goals for the unit.} First, \gfcrii{it} needs to be able to issue multiple memory accesses concurrently.
This is because (1)~we are designing the copy engine for an
arbitrarily-sized memory region (e.g., a column), which is often larger than 
the memory access granularity per vault (\sgcri{(8--256 B}) in an HMC-like \sgiii{memory}; and (2)~we 
want to fully exploit the internal bandwidth of 3D-stacked memory. 
Second, \sgiii{when a read for a copy completes, the accelerator
should immediately initiate the write.}}

\am{\gfdel{We design}\sgiii{Our} copy unit (\gfcri{Fig.}~\ref{fig:update-shipping-hw}\gfcrii{, right}) satisfies \sgiii{both} design goals. To issue multiple memory accesses concurrently,
we leverage the observation that these memory accesses are independent. We
use multiple \sgiii{fetch \gfcrii{(\incircledd{i} in Fig.~\ref{fig:update-shipping-hw})} and writeback \gfcrii{(\incircledd{ii})} units, which read from or write to
source/destination regions in parallel.} To satisfy the second design goal, we need to track outstanding reads, as
 they may come back from memory \sgiii{out of order}. 
Similar to \sgcri{prior work on accelerating \texttt{memcpy}}~\cite{memcopy-accelerator},
\sgiii{we use a \emph{tracking buffer} \gfcrii{(\incircledd{iii})} in our copy unit.  The buffer allocates} an entry for each read issued to memory. \gfdel{,
where a}An entry contains a memory address and a ready bit.
Once a read completes, we find its corresponding entry in \sgiii{the buffer} and
set its ready bit \sgiii{to trigger the write}.}

\am{\sgiii{We find that the buffer lookup limits the performance of the copy unit, as 
each lookup results in a full buffer scan,} and multiple fetch units \sgiii{perform
lookups concurrently (generating high contention)}. To \sgiii{alleviate this}, we design a hash index \gfcrii{(\incircledd{iv})} based on
the memory address to determine the location of a read in the buffer. \sgiii{We use a 
similar design as the hash lookup unit in our update \gfcri{gathering and} shipping unit \gfcrii{(\cref{sec:proposal:update-propagation:update-shipping})}.}}

\gfcrii{Our copy unit can be further accelerated \gfcriii{by using mechanisms that} provide fast in-DRAM \gfcriii{data} copy support~\gfcriii{\cite{seshadri2013rowclone, chang.hpca16, rezaei2020nom, olgun2021pidram, gao2020computedram, ambit,seshadri2019dram}}.}

\section{Analytical Engine}
\label{sec:proposal:analytic-engine}

The analytical execution engine\gfcri{, \gfcrii{whose} hardware design is illustrated in Fig.~\ref{fig:high-level-hw},} performs the analytical queries. \gfcrii{Our analytical engine consists of four simple programmable in-order PIM cores, placed within a vault of our 3D-stacked memory (i.e., \gfcriii{a} total of 64 PIM cores across the entire analytical island).} 
When a\gfcrii{n} \gfcri{analytical} query arrives, the \gfcri{analytical} engine parses the query and generates \rev{an algebraic} query plan
consisting of \gfcrii{\emph{physical operators}} (e.g., scan, filter, join).
\rev{In the query plan, operators are arranged in a tree where data flows from the bottom nodes
(leaves) toward the root, and the result of the query is stored in the root. The analytical execution 
engine employs the top-down Volcano (Iterator) execution model~\cite{volcano, volcano2} to traverse
 the tree and execute operators while respecting dependencies between operators. Analytical queries
 typically exhibit a high degree of both intra- and inter-query parallelism~\cite{qtm,morsel,scaling-up-c-store-numa}. To exploit this, the \gfcri{analytical} engine decomposes a\gfcri{n analytical} query into multiple tasks, each \gfcrii{of which is} a sequence of one or more operators.}
The \gfcri{analytical} engine \am{(task scheduler)} then schedules the tasks \gfcrii{with the goal of}
execut\gfcrii{ing} multiple independent tasks in parallel.

Efficient analytical query execution strongly depends on (1)~data placement,
(2)~the task scheduling policy, and (3)~how each physical operator is executed. Like prior works~\cite{mondrian, JAFAR}, we find that the \rev{execution of} 
\info{defined above}physical operators of analytical queries significantly benefit from PIM. 
However, without a\gf{n} HTAP-aware and PIM-aware data placement strategy and task scheduler,
PIM logic for operators alone \emph{cannot} provide significant throughput improvements.

We design \sgii{a new} analytical execution engine based on the characteristics of our in-memory hardware.
As we discuss in \gfcri{\cref{sec:proposal}}, \titleShort uses 3D-stacked
memory~\gfcrii{\cite{hmcspec2, hbm, lee2016smla}} that contains multiple vaults.
Each vault
(1)~provides only a fraction (e.g., 8 GB/s) of the total bandwidth available
in a 3D-stacked memory;
(2)~has limited power and area budgets for PIM logic; and
(3)~can access its own data faster than it can access data stored in other vaults\gfcrii{,} which take place through a vault-to-vault interconnect \gfcrii{(e.g., as in~\gfcriii{\cite{tesseract,azarkhish2014logic,hadidi2018performance,poremba2017there,azarkhish2016logic,ESMC_DATE_2015}}}).
We take these limitations into account as we design our data placement mechanism
and task scheduler.

\subsection{Data Placement}
\label{sec:proposal:analytic-engine:placement}

\begin{figure}[ht]
    \centering
        \centering
        \includegraphics[width=0.88\linewidth]{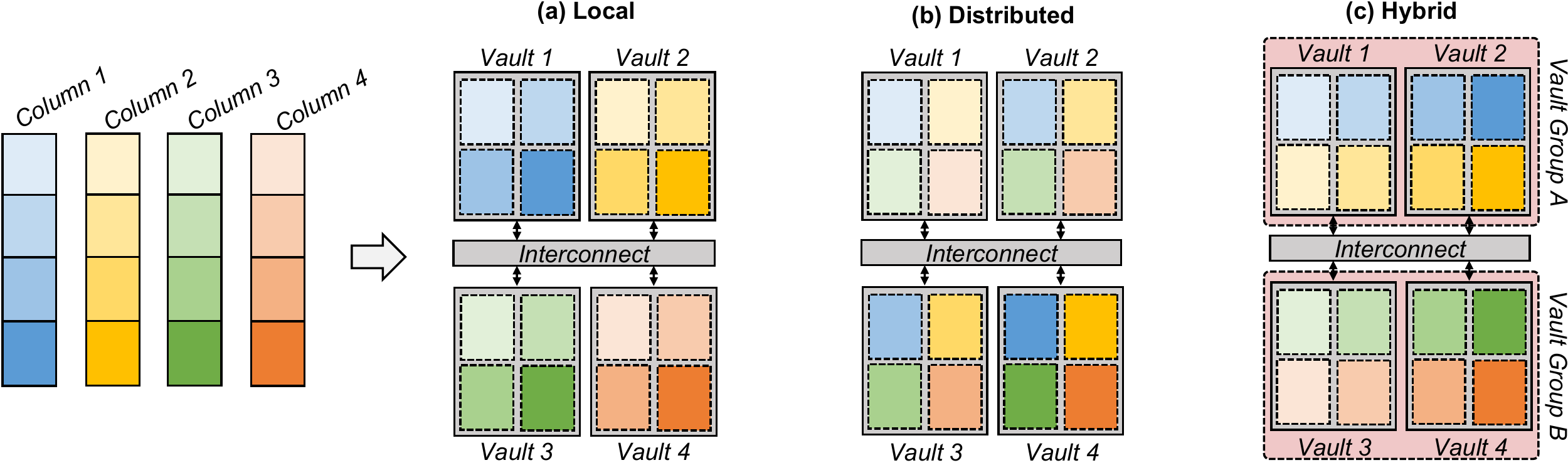}%
    \caption{\gfcrii{Three} data placement strategies.}
    \label{fig:data-placement}
\end{figure}

We evaluate three \gfdel{different }data placement strategies \gfcrii{(shown in Fig.~\ref{fig:data-placement})} for \titleShort.
Our analytical engine uses the DSM layout to store data, and
makes use of dictionary encoding~\cite{dict-compression} for column compression. 
Our three strategies affect which vaults the compressed DSM columns and dictionary are
stored in.

\gfrev{Strategy 1~\gfcriii{(\emph{Local})} stores the entire column (with dictionary) in one vault \gfcri{(Fig.~\ref{fig:data-placement}a)}, which improves analytical query throughput by making dictionary \gfcrii{lookups} and column accesses \gfcri{local to a single vault \gfcrii{(out of the 16 vaults we evaluate in~\cref{sec:eval})}. It also} simplifies the update application process\gfcri{,} since each vault has its own update application unit. However, this data placement strategy suffers from the limited area and memory bandwidth available in a single vault \gfcrii{since \gfcriii{an} analytical thread can only execute with the portion of the analytical replica placed at its assigned vault,} thus hurting throughput. Strategy 2~\gfcriii{ (\emph{Distributed})}  partitions \gfcrii{each} column across \gfcrii{\emph{all}} vaults in a \gfcri{memory} chip \gfcri{(Fig.~\ref{fig:data-placement}b)}, which allows \gfcri{the analytical engine} to exploit the entire internal bandwidth of the 3D-stacked memory and \gfcri{to} use all the available PIM logic to serve each \gfcri{analytical} query. However, this data placement strategy makes update application challenging due to gather-scatter operations \gfcrii{that span} all vaults, which reduces throughput.}

\gfrev{Strategy 3 \gfcriii{(\emph{Hybrid}),}  which \titleShort employs for data placement, overcomes the challenges of \gfcriii{\emph{Local}} and \gfcriii{\emph{Distributed}} by partitioning \gfcrii{a set of} columns across a \gfcri{\emph{vault group}} \gfcri{(Fig.~\ref{fig:data-placement}c)}. Each vault group consists of a fixed number of vaults, \gfcrii{each of which} holds a portion of the column. A group with \emph{v} vaults provides \emph{v} times the memory bandwidth and \sgcriii{\emph{v} times the PIM logic} power/area budget \sgcriii{for a column compared to \emph{Local}}.} \sgii{The number of vaults per group is critical for efficiency:} too many vaults can complicate the update application process, while not enough vaults can degrade throughput. We empirically find that four vaults per group strikes a good balance \gfcrii{(in a 3D-stacked memory with 16 vaults)}.

\gfcriii{The \emph{Hybrid} data placement strategy} still needs to perform \gfcrii{inter-vault} accesses within each vault group. To \sgii{overcome this}, we leverage an observation from prior work~\cite{update-c-store} that the majority of columns have only a limited (up to 32) number of distinct values. This means that the entire dictionary incurs negligible storage overhead (\textasciitilde 2 KB). \sgii{To avoid \gfcrii{inter-vault} dictionary accesses during update application, \gfcriii{\emph{Hybrid}}
keeps a copy of the dictionary in each vault.}
Such an approach is \sgii{significantly costlier} \gfcrii{if employed} under \gfcriii{\emph{Distributed}}\gfcrii{, as each vault \gfcriii{would} replicate \gfcriii{every} other \gfcriii{vault's} dictionary, generating large storage overheads due to \sgcriii{the vault count (e.g., 16)}.}

\subsection{\gfcrii{Task} Scheduler}
\label{sec:proposal:analytic-engine:scheduler}

\sgii{\titleShort's} task scheduler plays a key role in 
(1)~exploiting inter- and intra-query 
parallelism, and (2)~efficiently utilizing hardware resources. For each query, the \gfcrii{task} scheduler 
(1)~decides how many tasks to create, 
(2)~finds how to map these tasks to the available resources (\emph{PIM \gfcrii{cores}}), and 
(3)~guarantees that dependent tasks are executed in order. \rev{We first design a \gfcrii{\emph{basic task scheduler heuristic}} that generates tasks (statically at compile time)
by disassembling the operators of the query plan into operator instances (i.e., an
 invocation of a physical operator on some subset of the input tuples) based on (1)~which vault groups the input tuples reside in; and (2)~the number of
 available PIM threads in each vault group,
 which determines the number of tasks generated. The \gfcrii{basic task} scheduler \gfcrii{heuristic} inserts tasks into a global work queue
in an order that preserves dependencies between operators,
monitors the progress of PIM threads, and
assigns each task to a free thread (push-based assignment~\gfcrii{\cite{eager1986comparison}}). 

\gfdel{However, we find that this heuristic is not optimized for PIM, and leads to sub-optimal performance due to three reasons. First, the heuristic requires a dedicated
 runtime component to monitor and assign tasks.
The runtime component must be executed on a general-purpose PIM core,
either requiring another core (difficult given limited area/power budgets) or
preempting a PIM thread on an existing core (which hurts performance). Second, the heuristic's static mapping is limited to using only the resources
available within a single vault group, which can lead to performance issues
for queries that operate on very large columns. Third, this heuristic is vulnerable to load imbalance, as
some PIM threads might finish their tasks sooner and wait idly for straggling threads. }

We \gfcrii{design a\gfcriii{n} \emph{optimized task scheduler heuristic} by} \gfrev{\gfcrii{employing} three optimization\gfcrii{s} on top of our} \gfrev{basic} \gfcrii{task scheduler} heuristic} \gfrev{so \gfcrii{our task scheduler} can better fit our PIM system.} First, we design a pull-based task assignment strategy~\gfcrii{\cite{eager1986comparison}}, where PIM threads pull tasks from the task queue at runtime. This eliminates the need for a
 runtime component \gfdel{(first challenge) }and allows PIM thread\gfcrii{s} to 
 dynamically balance \gfcrii{their loads}\gfdel{ (third challenge)}. \gfdel{To this end, w}\gfrev{W}e introduce a local task queue for each vault group. Each PIM thread \gfcrii{examines} its
  own local task queue to retrieve its next task. Second, we optimize the heuristic to 
  allow for finer-grained tasks. 
 Instead of mapping tasks statically, we partition input tuples into fixed-size segments (i.e., 1000 tuples) and
  create an operator instance for each partition. The \gfcrii{task}
  scheduler then generates tasks for these operator instances
 and inserts them into \gfcrii{their} corresponding task queues (where those tuple segments reside).
 \gfcrii{A large} number of tasks increases opportunities for load balancing. \gfrev{Third}, we \gfdel{optimize the heuristic to} allow a PIM thread to steal tasks
 from a remote vault if its local queue is empty. This enables us to potentially use all available
 PIM threads to execute tasks, regardless of the data \gfcrii{placement}\gfdel{ (second challenge)}. Each PIM
 thread first attempts to steal tasks from other PIM threads in its own vault group, because the
 thread already has a local copy of the full dictionary in its vault, and needs \gfcrii{inter-}vault
 accesses only for the column partition. If there is no task to steal in its vault group, the PIM thread
 attempts to steal a task from a remote vault group.

\subsection{Hardware Design}
\label{sec:proposal:analytic-engine:hardware}

\gfrev{Our analytical engines leverage the design of our new data placement strategy and \gfcrii{task} scheduler to expose \info{Mostly intra.}intra-query parallelism and the available vault bandwidth to PIM cores.} We add four \gfrev{simple programmable in-order PIM cores~\gfcrii{\cite{Mingyu:PACT,mondrian,tesseract}}}
to each vault\gfdel{, where the cores are similar to those in prior work~\cite{Mingyu:PACT,mondrian}}.
\change{We run} a PIM thread on each core, and \change{we use} these cores to execute
the \gfcrii{task} scheduler and other parts of the analytical engine (e.g., \info{Discussed at the beginning of Sec. 7}\gfcrii{the} query parser).

\rev{We find that our optimized \gfcrii{scheduler} heuristic significantly increases data sharing
    between PIM threads. This is because within each vault group, all 16 PIM threads
        access the same local task queue, and must synchronize their accesses. The problem worsens when other PIM threads attempt to steal tasks from remote vault groups, especially for highly-skewed workloads. To avoid excessive accesses to DRAM and let PIM
        threads share data efficiently, we implement a simple fine-grained coherence technique (as in~\gfcriv{\cite{conda,amiraliphd}}), which uses
        a local PIM-side directory in the logic layer to implement a low-overhead coherence protocol.}

\section{\gfrev{Integrating \titleShort in the Cloud}}
\label{sec:integrating:cloud}

\gfrev{\gfcrii{We} discuss the applicability of our proposal to cloud environments. First, we discuss the benefits of integrating \titleShort in the cloud. Second, we discuss alternative cloud-based hardware solutions for HTAP systems.}  

\gfrev{\textbf{Integrating \titleShort in the Cloud.} We believe that cloud providers would be willing to add extra ASICs into their environment as long as doing so provides potential improvements to a wide enough range of cloud workloads. Several cloud providers already integrate ASIC accelerators in their datacenters. For example, Google Cloud, Microsoft Azure and Amazon AWS use dedicated hardware for neural network inference (Google Cloud TPU~\gfcrii{\cite{CloudTpu58,jouppi2021ten}}, Habana Goya~\cite{medina2020habana}), neural network training (AWS Trainium~\cite{AWSTrain0}, Habana Gaudi~\cite{medina2020habana}), video transcoding (Google VCU~\cite{ranganathan2021warehouse}), and compression/encryption/data authentication (Microsoft's Project Corsica~\cite{Improved12}). Recently, different commercial PIM designs~\gfcrii{\cite{kim2021aquabolt,devaux2019true, lee2022isscc,kwon202125,lee2021hardware,ke2021near,niu2022184qps,gomez2021benchmarking,gomez2021benchmarkingcut}}, which target large cloud systems, have been proposed.\footnotemark[1] Since Polynesia’s hardware can provide performance benefits \gfcrii{(\cref{sec:eval:analysis})} across a wide range of applications that depend on large amounts of data and analytics, we believe it is an attractive architecture for cloud providers. Polynesia’s energy savings \gfcrii{(\cref{sec:eval:energy})} are also attractive to a cloud environment, as it can significantly lower operating costs. As prior works show~\cite{magaki2016asic,ranganathan2021warehouse}, ASIC accelerators are a viable solution for cloud environments, depending on the scale of the computation. Even though we cannot accurately predict the price of integrating Polynesia into a cloud system due to unknown parameters (e.g., non recurring engineering expenses), we believe it would be a beneficial solution for cloud systems due to the \gfcrii{widespread use} of the database applications it targets.  }

\gfrev{\textbf{Alternative Hardware Solutions.} \gfcrii{A straightforward alternative solution to improve performance for a cloud HTAP system is to scale up hardware resources (especially core count).} However, \gfcrii{as we demonstrate in this paper,} several key bottlenecks in HTAP systems are not compute-bound but are instead memory-bound and cannot benefit simply \gfcrii{from} more cores (as memory bandwidth\gfcrii{, latency, and data movement remain as bottlenecks with more cores}). PIM hardware can overcome the memory bandwidth bottlenecks by tapping into the significantly higher internal memory bandwidth of 3D-stacked memories and \gfcrii{also} improving \gfcrii{latency and} energy efficiency \gfcrii{with reduced data movement}. Therefore, our custom hardware has distinct benefits unachievable by additional cores.}

\section{Methodology}
\label{sec:methodology}

We use and heavily extend state-of-the-art transactional and analytical engines to 
implement various single- and multiple-instance HTAP configurations.
We use DBx1000~\cite{dbx1000,dbx1000-github} as the starting point for
our transactional engine, and we implement an in-house analytical engine
similar to C-store~\cite{c-store}.
Our analytical engine supports key physical
operators for relational analytical queries (select, filter, aggregate and join), and
supports both NSM and DSM layouts, and dictionary encoding~\cite{dict-compression,compression-c-store,scaling-up-c-store-numa}.
For consistency, we implement both snapshotting (similar to software 
snapshotting~\cite{software-snapshotting}, \sgii{with snapshots taken 
only when dirty data exists}) and MVCC (\info{adopted.}adopted from DBx1000~\cite{dbx1000}). 

Our baseline single-instance HTAP system
stores the single data replica in main memory.
Each transactional query randomly performs reads or writes on a few randomly-chosen tuples from a randomly-chosen
table. Each analytical query uses select and join on randomly-chosen tables and columns. 
Our baseline multiple-instance HTAP system models a similar system as our single-instance baseline,
but provides the transactional and analytical engines with separate replicas
(using the NSM layout for transactions, and DSM with dictionary encoding for analytical queries).
Across all baselines, we have 4 transactional
and 4 analytical worker threads. 

We simulate \titleShort using gem5~\cite{GEM5},
integrated with DRAMSim2~\cite{DRAMSim2} to model 
an HMC-like 3D-stacked DRAM~\cite{hmcspec2}.
Table~\ref{tbl:config} shows our system configuration.
For the analytical island, 
each vault of our 3D-stacked memory contains
four PIM cores and three fixed-function accelerators (update \gfcri{gathering and} shipping unit,
update application unit, copy unit).
For the PIM core, we model a core
similar to the ARM Cortex-A7~\cite{cortex-a7}\gfcrii{, as in prior \sgcriii{works}~\cite{syncron,picorel2017near}}.

\begin{table}[ht]
    \caption{Evaluated system configuration.}
    \label{tbl:config}%
    \small
    \renewcommand{\arraystretch}{0.5}
   \resizebox{\columnwidth}{!}{
    \begin{tabular}{ll}
        \toprule
        \emph{Processor} &  
        4 OoO cores, \change{each with 2 HW threads}, 8-wide issue; \\ 
        \emph{(Transactional} &     
        \emph{L1 I/D Caches}: 64~kB private, 4-way assoc.; \emph{L2 Cache:} \\
	\emph{Island)} & 8~MB shared, 8-way assoc.; \emph{Coherence}: MESI~\cite{papamarcos1984low} \\
        \midrule
        \emph{PIM Core\gfcrii{s}} & 
           4 in-order cores per vault, 2-wide issue, \\ 
        \gfcrii{\emph{(Analytical}} & \emph{L1 I/D Caches}: 32~kB private, 4-way assoc. \\
        \gfcrii{\emph{Engine)}}  & \gfcrii{\emph{Coherence}: MESI}~\cite{papamarcos1984low} \\
        \midrule
        \emph{3D-Stacked} & 
              4 GB cube, 16 vaults per cube; \emph{Internal Bandwidth:} \\
              \emph{Memory} &  256 GB/s; \emph{Off-Chip Channel Bandwidth:} 32 GB/s \\
        \bottomrule
    \end{tabular}%
    }
\end{table}

\gf{We \gfcriii{model the} in-order cores and specialized accelerator\gfcriv{s} in gem5 using a methodology \gfcrii{similar} to\gfdel{ prior works}~\gfcriv{\cite{google-pim, conda, LazyPIM,amiraliphd}}. Similar to prior works~\gfcriv{\cite{conda, LazyPIM,amiraliphd}}, we implement a simple fine-grained coherence technique between PIM cores\gfdel{(we discussed the reasoning for having coherence between PIM cores in \gfcri{\cref{sec:proposal:analytic-engine:hardware}}. The coherence mechanism uses a local PIM-side directory in the logic layer to implement a low-overhead coherence protocol}. Updates between islands are propagated using our update propagation mechanism and shared memory. The updates are stored in shared memory. \gfdel{The update propagation mechanism handles the communication as we describe in \gfcri{\cref{sec:proposal:update-propagation}}.} To allow coordination, ordering, and synchronization between different parts of islands, we need to provide coherence between CPU cores and PIM logic. We employ a simple fine-grained coherence technique \gfcriii{(MESI~\cite{papamarcos1984low})}, which uses a local PIM-side directory~\gfcriii{\cite{censier1978new}} in the logic layer to maintain coherence between PIM cores and to enable low-overhead fine-grained coherence between PIM logic and the CPUs. \gfdel{The CPU-side directory acts as the main coherence point for the system, interfacing with both the processor caches and the PIM-side directory.}}

\sgcriii{We open-source \titleShort and the complete source code of our evaluation~\cite{polynesia.github}.}

\gfrev{\textbf{Experimental Setup Validation}. We validate our experimental setup, including our workloads and gem5-based simulation, in two ways.  First, to implement our HTAP system, we adopt \emph{prior} transactional (DBx1000) and analytical (C-store) engines. Similarly, we \gfcriii{tailor our} consistency models \gfcriii{based on} prior works~\cite{software-snapshotting, dbx1000}. We validate the correctness of each \gfcrii{model} modification we ma\gfcrii{k}e by comparing the outputs \emph{after} modifications with the outputs \emph{before} modifications. Second, to model the hardware components of our system, we use an already validated simulator (i.e., gem5).  While we expect to see similar results when Polynesia is implemented on top of high-end HTAP systems, it is currently very challenging for us to confirm that in real hardware \gfrev{since Polynesia requires hardware modifications that are costly to realize or prototype using available tools and frameworks}.}\footnote{\gfrev{While PiDRAM~\cite{olgun2021pidram} and MEG~\cite{zhang2020meg} are \gfcrii{potential promising} FPGA-based platforms to emulate the functional correctness of proposed PIM designs, they are not useful for our studies because (1)~neither can model the performance and energy usage of PIM, (2)~PiDRAM does not support the 3D-stacked memories used by Polynesia, and (3)~the MEG framework is not available at the time of writing.\gfdel{Polynesia employs 3D-stacked memories and PIM as part of its design. Modeling PIM in 3D-stacked memories in common hardware prototyping platforms (e.g., FPGAs) is challenging due to a lack of tools and frameworks that accurately prototype state-of-the-art PIM designs. Two prior works, PiDRAM~\cite{olgun2021pidram} and MEG~\cite{zhang2020meg}, propose FPGA-based frameworks to study the functional behavior of PIM components. However, they are not helpful for our studies since neither accurately captures the performance and energy of PIM. Furthermore, (i) PiDRAM only supports DDR3 devices, which have a different interface and much lower bandwidth than Polynesia’s 3D-stacked memory, and (ii) the source code of MEG is not available.}} }

\subsection{\gfrev{Simplifying Assumptions About Realistic HTAP}}
\label{sec:methodology:assumptions}
\gfrev{We had to make simplifying assumptions to capture most of the key properties of a real HTAP system in an architectural simulator. We list such assumptions next.}

\gfrev{\textbf{Scheduling Transactional and Analytical Queries.} We assume that the partitioning between transaction\gfcrii{al} and analytical queries is made by the DBMS. We do not propose a specific dynamic scheduling mechanism to identify the query type and map the query to the appropriate island. We believe that such a dynamic scheduling mechanism is orthogonal to our work, and can be an extension of existing mechanisms~\cite{task-sched-anal-txn,scaling-up-htap,raza2020adaptive}.}

\gfrev{\textbf{Modern Transactional and Analytical Engines.} Our goal in this paper is to (1)~provide insights about the major challenges in HTAP systems, (2)~propose a framework that can address these challenges, and (3)~evaluate such a framework as a case study. To achieve this goal, we employ transactional and analytical models that may not necessarily represent the most recent implementations of OLTP and OLAP workloads, but are enough to demonstrate the challenges and drawbacks that most HTAP systems face due to data movement overheads. The transactional and analytical models we use for our reference implementation of \titleShort have the following advantages. First, our transactional engine (DBX1000) is a simple, in-memory DBMS that provides several concurrency control models, including MVCC and the optimistic concurrency control model, a similar but improved version of the concurrency control algorithm Microsoft's Hekaton~\cite{diaconu2013hekaton} employs. Many works~\cite{yu2016tictoc,xia2020taurus,yu2018sundial,PICA} build on top of DBX1000, proposing both hardware~\cite{PICA} and software~\cite{yu2016tictoc,xia2020taurus,yu2018sundial} optimizations. Second, our analytical engine is a column-based analytical engine adopted from C-store, which employs a Volcano-style processing model~\cite{graefe1993volcano}. \gfrev{While modern analytical engines may employ more efficient processing models  (e.g., based on vectorization~\cite{boncz2005monetdb} or pushing tuples~\cite{neumann2014compiling}), our Volcano-style processing model is still present in many relational DBMS\gfcrii{s}~\cite{qem}. To conclude, we use such transactional and analytical models since they are enough to reach our goals. We believe that the challenges we identify in HTAP systems are also present in systems with more modern transactional and analytical engines\gfcrii{, but we leave detailed studies of large-scale systems to future works}.} }

\gfrev{\textbf{Assumptions About the Encodings/Formats.} Our analytical
engine uses dictionary encoding, similarly to prior works~\cite{dict-compression,compression-c-store}. However, dictionary encoding might not be beneficial to a particular real system if there are not enough common values across columns to employ dictionary compression.}

\section{Evaluation}
\label{sec:eval}

\gf{We demonstrate the advantages of \titleShort by evaluating (i) its end-to-end \gfcrii{performance} benefit\gfcrii{s} compared against state-of-the-art HTAP systems using both synthetic and real\gfcri{-}world queries; (ii) the individual performance of the three major components of our proposal (i.e., \gfcrii{our} update propagation \gfcrii{technique}, consistency \gfcriii{mechanism}, and analytical
engine); (iii) how  \titleShort performs as the dataset size grows; and (iv) the energy savings \titleShort provides.}

\sgdel{We first evaluate the three major components of our proposal:
(1)~update propagation, (2)~our consistency mechanism, and (3)~our analytical
engine. We then perform an end-to-end system evaluation.}

\subsection{\gfcriv{End-to-End System Performance Analysis}}
\label{sec:eval:analysis}

\begin{figure}[ht]
    \centering
        \centering
        \includegraphics[width=\linewidth]{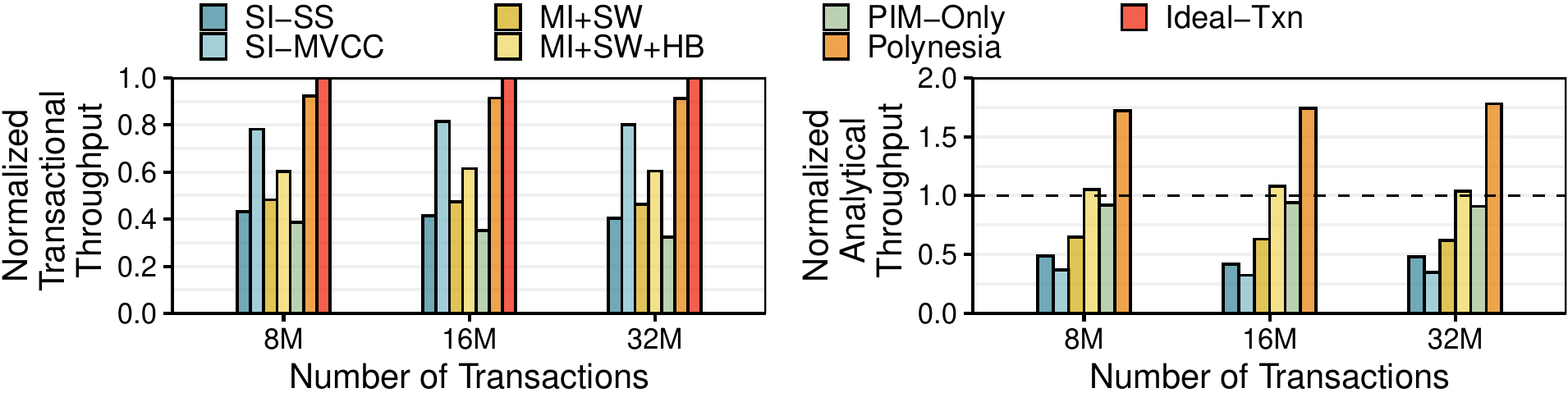}%
    \caption{Normalized transactional (left) \change{and analytical (right)} throughput for end-to-end HTAP systems.}
    \label{fig:eval-end-to-end}
\end{figure}

\gfcri{Fig.}~\ref{fig:eval-end-to-end} \change{(left)} shows the transactional throughput \change{\gfcrii{of} \rev{six} DBMSs:
(1)~Single-Instance-Snapshot (\rev{\emph{SI-SS}}; \gf{modeled after the Hyper HTAP system~\cite{hyper} with software snapshotting~\cite{software-snapshotting}});
(2)~Single-Instance-MVCC (\emph{SI-MVCC}; \gf{modeled similar to AnkerDB~\cite{ankerdb}});
(3)~\emph{MI+SW}, an improved version of
 Multiple-Instance, \gf{modeled similar to Batch-DB~\cite{batchdb}} \gfcri{and} includ\gfcrii{ing} all of our software optimizations for \titleShort (except those
 specifically targeted for PIM);
(4)~\emph{MI+SW+HB}, a hypothetical version of 
\emph{MI+SW} with 8$\times$ \gfcri{its} \gfcri{main memory} bandwidth (256 GB/s), equal to the
 internal \gfcri{memory} bandwidth of \gfcrii{a commercially available 3D-stacked memory (}HBM \gfcrii{2.0}~\gf{\cite{hbm}}\gfcrii{)}; 
 \rev{(5)~\emph{PIM-Only}, a hypothetical version of \emph{MI+SW} \gfcri{that} uses general-purpose PIM cores to run both transactional and analytical workloads; and
(6)}~\emph{\titleShort}, our full hardware--software proposal.} \gf{Each one of these baselines (i)~isolates one of our new components and shows how much benefit we get out of them; (ii)~is modeled after state-of-the-art (software-only) HTAP systems.} \gf{Note that we do not compare \titleShort against hardware-based HTAP systems since no prior work has proposed to use \gfcri{tailored} hardware accelerators for HTAP.\gfdel{ This is the \emph{first} work that uses software/hardware co-design to achieve all desired properties of an HTAP system.}} We normalize throughput to an ideal transaction-only DBMS (\emph{Ideal-Txn}) \change{for \gfcri{three} transaction count\gfcri{s}}. \gfrev{\emph{Ideal-Txn} indicates the peak transactional throughput if we run the transactional workload in isolation.}

\gf{We make five observations from \gfcrii{Fig.~\ref{fig:eval-end-to-end} (left)}.}
\gfcri{First,  \titleShort improves transactional throughput by 51.0\% over \emph{MI+SW+HB}, and by 14.6\% over \emph{SI-MVCC}, while achieving 91.6\% of the transactional throughput of \sgcri{\emph{Ideal-Txn}}, on average across the three transaction \sgcri{counts. \titleShort's} higher transactional throughput stems from (1)~using custom PIM logic for analytical queries and update propagation and consistency mechanisms that reduce resource contention significantly, and (2)~reducing off-chip memory bandwidth contention by decreasing data movement.}
\gf{Second,}\change{ of the single-instance DBMSs, \emph{SI-MVCC} performs best, \gfcri{achieving 8}0.0\% of the throughput of \emph{Ideal-Txn}\gfcri{, on average across the three transaction counts}.
Its use of MVCC over snapshotting overcomes the high performance penalties incurred by
\emph{SI-SS}.
\gfcri{Third, \emph{SI-MVCC} significantly outperforms the two software-only multiple-instance DBMSs (\emph{MI+SW} and \emph{MI+SW+HB}), \gfcrii{even though the latter two} enabl\gfcrii{e} software optimizations \sgcri{and \gfcrii{greatly} increas\gfcrii{e the} memory bandwidth}. Such performance degradation is due to \sgcri{the lack of performance isolation in both \emph{MI+SW} and \emph{MI+SW+HB}, and} in the case of \emph{MI+SW}, the \gfcrii{large} data movement overhead of update propagation. This demonstrates that we \emph{cannot} achieve the three desired properties of an HTAP system without co-designing software with hardware.}}
\gfcri{Fourth, \change{\emph{MI+SW+HB}} cannot mitigate data movement and contention on shared resources, even with \gfcrii{8$\times$ the} main memory bandwidth. As a result, \emph{MI+SW+HB} transactional throughput is \gfcrii{58.8\% of that of} \emph{Ideal-Txn}.} \gf{Fifth,} \rev{\gfcrii{\emph{PIM-Only}} significantly hurts transactional throughput (by 67.6\% vs. \emph{Ideal-Txn}),
 and performs 7.6\% worse than \emph{SI-SS}. \gfcrii{This happens because transactional queries have cache-friendly memory access patterns~\info{Correct}\gfcriv{\cite{conda, LazyPIM,amiraliphd}}, thus benefiting from a deep cache hierarchy, which is not present in \emph{PIM-Only}.}}

\change{\gfcri{Fig.}~\ref{fig:eval-end-to-end} (right) shows the analytical throughput \gfcrii{of the} \gf{six} DBMSs. We normalize \gfcri{the analytical} throughput at each transaction count to a baseline where analytical queries are running alone on \gfcri{the} system \gfcrii{(\emph{Base-Anl})}. \gf{We make four observations.} 
\gfcri{First, \titleShort \emph{improves} analytical throughput over \gfcrii{\emph{Base-Anl}} by 63.8\%, on average across the three transaction counts. \sgcri{The performance benefits of \titleShort} come from eliminating data movement, \gfcrii{reducing the} latency \gfcrii{of} \sgcri{memory} accesses, and using custom logic for update propagation and consistency. 
Second, while \emph{SI-MVCC} is the best software-only DBMS when considering transactional throughput, it \emph{degrades} analytical throughput by 63.2\% compared to \gfcrii{\emph{Base-Anl}} due to its lack of workload-specific optimizations and \gfcrii{pointer-chasing intensive} consistency mechanism (MVCC\gfcrii{; see~\cref{sec:bkgd:single}}). 
Third, \emph{MI+SW+HB} improves analytical throughput by 41.2\% over \emph{MI+SW}. However, it still suffers an analytical throughput loss of 35.5\% compared to \gfcrii{\emph{Base-Anl}}, on average, even though  \emph{MI+SW+HB} has \gfcrii{8$\times$ the} main memory bandwidth \gfcrii{of} \emph{MI+SW}. 
Fourth, the analytical throughput of \emph{PIM-Only} is 11.4\% lower than that of \emph{MI+SW+HB}, on average, as \emph{PIM-Only} suffers from resource contention caused by co-running transactional and analytical queries.}

} 
\gfcri{We conclude that \titleShort's \sgcri{island-based} hardware-software co-design leads to significant perform\gfcrii{ance} benefits compared to \gfcrii{all evaluated} software-only implementations.} \sgii{\gfrev{A}veraged across all transaction counts in \gfcri{Fig.}~\ref{fig:eval-end-to-end}, 
\titleShort has \gfcrii{both} a higher transactional throughput
(2.20$\times$ over \emph{SI-SS}, 1.15$\times$ over \emph{SI-MVCC}, and 1.94$\times$ over \emph{MI+SW};
mean of 1.70$\times$) \sgcri{\emph{and}} a higher analytical throughput
(3.78$\times$ over \emph{SI-SS}, 5.04$\times$ over \emph{SI-MVCC}, and 2.76$\times$ over \emph{MI+SW};
mean of 3.74$\times$).}

\paratitle{Real Workload Analysis} To model more complex queries, we evaluate \titleShort using a mixed workload 
from TPC-C~\cite{TPC-C} (for our transactional workload) and TPC-H~\cite{TPC-H} (for our analytical workload). TPC-C's 
schema includes nine relations (tables) that simulate an order processing application. We simulate two
 transaction types defined in TPC-C, \emph{Payment} and \emph{New order}, which together account for 88\% of the TPC-C
 workload~\cite{dbx1000} \gfrev{and touch all nine tables defined by TPC-C}. We vary the number of warehouses from 1 to 4, and we assume that our
 transactional workload includes an equal number of transactions from both \emph{Payment} and \emph{New order}. \gfrev{\sgrevi{For TPC-H, we use three of its queries in our experiments, each of which displays different behavior and  operates over six TPC-H tables (out of the eight total tables in the TPC-H schema). The six tables are \emph{LINEITEM}, \emph{PART}, \emph{SUPPLIER}, \emph{PARTSUPP}, \emph{ORDERS}, and \emph{NATION}, with a cardinality (i.e., number of rows) of  6M, 200K, 10K, 800K, 1.5M, and 25, respectively.}}  
\gfrev{The three TPC-H queries we evaluate are:
(i)~Query 1 (\emph{Q1}), an aggregation-heavy query~\cite{boncz2013tpc} that generates a pricing summary report over the \emph{LINEITEM} table;
(ii)~Query 6 (\emph{Q6}), a selection-heavy query~\cite{boncz2013tpc} that computes a forecast revenue change over the \emph{LINEITEM} table}\gfrev{; and
(iii)~Query 9 (\emph{Q9}), a join-heavy query~\cite{boncz2013tpc} that measures the profit for a given product type over all six tables.}

We evaluate the transactional and analytical throughput for \titleShort and for three baselines: (1)~\emph{SI-SS}, (2)~\emph{SI-MVCC}, (3)~\emph{MI+SW}. 
We find that, averaged across all warehouse counts, \titleShort has a higher \sgrevi{throughput \gfcrii{than} all three baselines. More specifically, \titleShort achieves both a higher transactional throughput} (2.31$\times$ over \emph{SI-SS}, 1.19$\times$ over \emph{SI-MVCC}, and 1.78$\times$ over \emph{MI+SW}; mean of 1.76$\times$) and a higher analytical throughput \gfrev{for 
\emph{Q1} (2.84$\times$ over \emph{SI-SS}, 4.12$\times$ over \emph{SI-MVCC}, and 2.4$\times$ over \emph{MI+SW}; mean of 3.04$\times$), 
for \emph{Q6}} (3.41$\times$ over \emph{SI-SS}, 4.85$\times$ over \emph{SI-MVCC}, and 2.2$\times$ over \emph{MI+SW}; mean of 3.48$\times$)\gfrev{, and 
for \emph{Q9} (3.67$\times$ over \emph{SI-SS}, 4.51$\times$ over \emph{SI-MVCC}, and 1.95$\times$ over \emph{MI+SW}; mean of 3.18$\times$)}.

\change{We conclude that \sgii{\titleShort's ability to meet all three
HTAP properties enables better transactional and analytical performance over all three \gfcrii{evaluated}
state-of-the-art systems.
\gfdel{In \gfcri{\cref{sec:eval:update-propagation}}, \gfcri{\cref{sec:eval:consistency}}, and
\gfcri{\cref{sec:eval:analytical}}, we study how each component of \titleShort
contributes to performance.}}}

\subsection{\gfcrii{Effect of} \gfcrii{the} Update Propagation \gfcriii{Technique}}
\label{sec:eval:update-propagation}

\begin{figure}[h]
    \centering
        \centering
        \includegraphics[width=\linewidth]{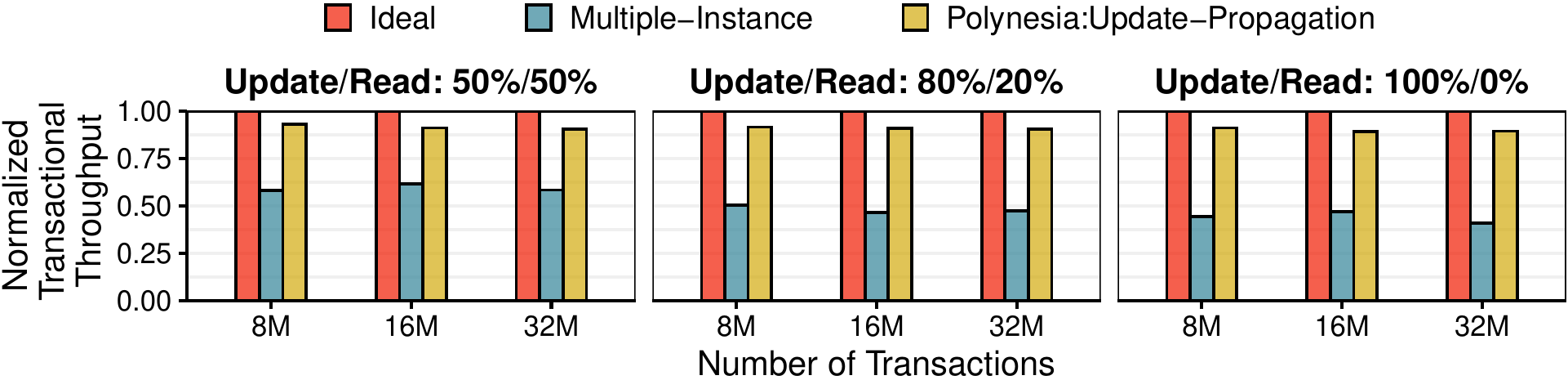}%
    \caption{Effect of update propagation mechanisms on transactional throughput.}
    \label{fig:eval-update-propagation}
\end{figure}

\gfcri{Fig.}~\ref{fig:eval-update-propagation} shows
 the transactional throughput for \titleShort's update propagation
 mechanism and \emph{Multiple-Instance}, normalized to a multiple-instance baseline with zero cost for update propagation
 (\emph{Ideal}). We assume \gfcrii{that} each \sgcri{thread of the analytical workload} executes 128 queries, and vary both the number of transactional
 queries per \sgcri{thread} and the
 transactional query \gfcri{update-to-read} ratio. 
To isolate the impact of different update propagation mechanisms,
we use a zero-cost consistency mechanism, and ensure
that the level of interference \gfcrii{between transactional and analytical threads} remains the same for all mechanisms. \gf{We make \gfcri{two} observations \gfcri{from \gfcrii{Fig.~\ref{fig:eval-update-propagation}}}.}

\gfcri{First, \titleShort's update propagation mechanism improves transactional throughput by 1.8$\times$ compared to \emph{Multiple-Instance}, and comes within 9.2\% of \emph{Ideal}, on average across all transaction counts and update-to-read ratios. The improvement comes from (1)~significantly reducing data movement by 
 offloading \gfcrii{the} update propagation process to PIM, (2)~freeing up CPUs from performing update propagation by using a specialized  hardware accelerator, and (3)~\gfcrii{co-designing} \gfcriii{the} hardware and software \gfcrii{for update propagation}. Overall, our mechanism reduces the latency of update propagation by 1.9$\times$ compared to \emph{Multiple-Instance} (not shown). }
\gfcri{Second,} we find that \emph{Multiple-Instance} degrades transactional throughput, on average, by 49.5\% compared to \emph{Ideal}, as it severely suffers from resource contention \gfcrii{(e.g., at shared caches and main memory)} and data movement cost. \gfcri{We observe that} 27.7\% of \emph{Multiple-Instance}'s transactional throughput degradation comes
from the update \gfcri{gathering and} shipping latencies associated with \gfcrii{(i)}~data movement and \gfcrii{(ii)}~merging updates from multiple transactional threads. The remaining 21.8\% transactional throughput degradation is due to the update application process,
where the major bottlenecks are column compression/decompression and dictionary reconstruction.

We conclude that \gfcri{\titleShort's} update propagation mechanism provides data freshness (i.e., low update latency) while maintaining high transactional
throughput (i.e., performance isolation).

\subsection{\gfcrii{Effect of the} Consistency Mechanism}
\label{sec:eval:consistency}

\begin{figure}[h]
    \centering
        \includegraphics[width=\linewidth]{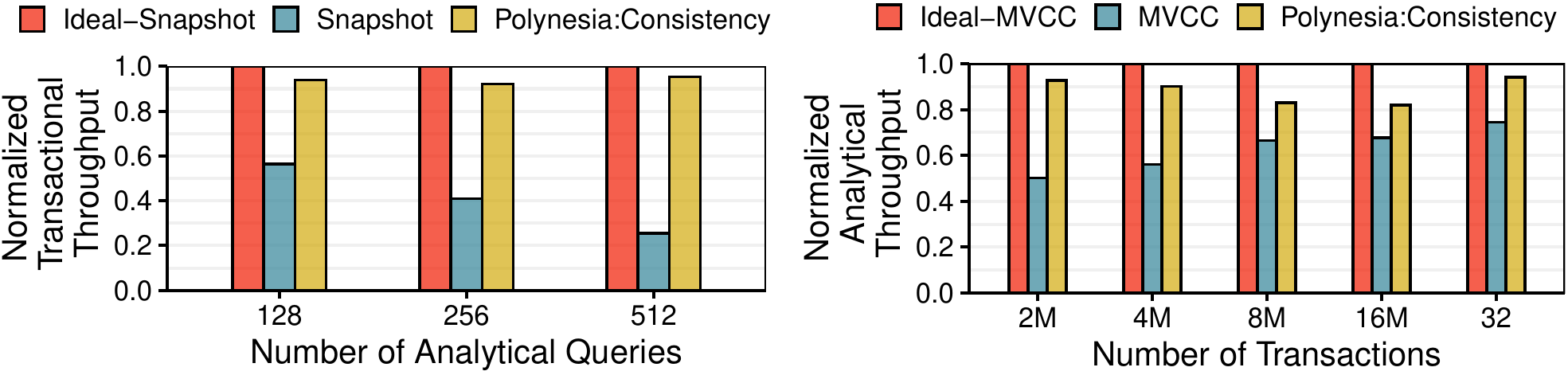}%
    \caption{Effect of consistency mechanisms on \gfcrii{transactional} (left) and \gfcrii{analytical} (right) throughput.}
    \label{fig:eval-mvcc}
\end{figure}

 \gfcri{Fig.}~\ref{fig:eval-mvcc} (\gfcriii{left}) shows
 the transactional throughput for \titleShort's consistency mechanism
and Single-Instance-Snapshot
 (\emph{Snapshot}), normalized to a single-instance baseline with zero-cost snapshotting (\emph{Ideal-Snapshot}). Each thread performs 1M transactional queries  \gfcri{and we vary the analytical query count}.
\gf{We make two observations. First,} \titleShort improves transactional throughput by 
2.2$\times$ over \emph{Snapshot}, and comes within 6.1\% of
 \emph{Ideal-Snapshot}, \gfcri{on average,} because it \sgmod{performs snapshotting}{snapshots} at a column
 granularity and leverages PIM \sgmod{to perform}{for} fast snapshotting. \gfcri{Second,} \emph{Snapshot} reduces transactional throughput, on average, by 59\% compared  to \emph{Ideal-Snapshot}. This is because of expensive \texttt{memcpy} operations needed to create  each snapshot, resulting in significant \gfcrii{memory bandwidth} contention. 

\gfcri{Fig.}~\ref{fig:eval-mvcc} (\gfcriii{right}) shows the analytical throughput of
 \titleShort's consistency mechanism and of Single-Instance-MVCC (\emph{MVCC}), normalized to a single-instance baseline with zero cost for MVCC (\gfrev{\emph{Ideal-MVCC}, a hypothetical
ideal baseline where MVCC operations incur zero delay during execution}) \gfcri{for five transaction counts}. We assume each \sgcri{thread of the analytical workload} executes 128 queries, and we vary the transactional query count per \sgcri{analytical workload} thread. 
 For a fair comparison, we implement our consistency 
 mechanism in a single-instance system. \gf{We make two observations. First,} \gfcri{\titleShort's} consistency mechanism improves analytical throughput by 
1.4$\times$ compared to \emph{MVCC}, and comes within 11.7\% of
 \emph{Ideal-MVCC}, \gfcri{on average, since \titleShort's consistency mechanism allows} analytical queries to \gfcri{avoid} scan\gfcri{ning} lengthy version chains when accessing
 each tuple. \gfcri{Second,} \sgdel{We find that }\emph{MVCC} degrades analytical throughput, on average, by 37.0\% compared to
 \emph{Ideal-MVCC}, as it forces each analytical query to traverse a lengthy version chain and perform expensive timestamp comparisons to locate the most recent version \gfcri{of the data}.  

We conclude that \gfcri{\titleShort's} consistency mechanism maintains consistency without compromising
performance isolation\gfcrii{, leading to high analytical and transactional throughput}.

\subsection{\gfcrii{Effect of the} Analytical Engine}
\label{sec:eval:analytical}

We study the effect of each of our data placement 
 strategies from \gfcri{\cref{sec:proposal:analytic-engine:placement}}: 
\gfcri{(i)}~\gfcriii{\emph{Local}};
\gfcri{(ii)}~\gfcriii{\emph{\gfcriii{Distributed}}};
\gfcri{(iii)}~our \emph{Hybrid} strategy, where all use the \info{defined in 7.2.}basic scheduler heuristic\gfcri{; and}
\gfcri{(iv)}~our hybrid strategy combined with our \gfcrii{optimized} \gfcrii{task} scheduler
 heuristic (labeled as \emph{Hybrid-\sgcri{Sched}}).
For these studies, \gfcri{analytical queries address the same column}.

\begin{figure}[ht]
    \centering
        \centering
        \includegraphics[width=\linewidth]{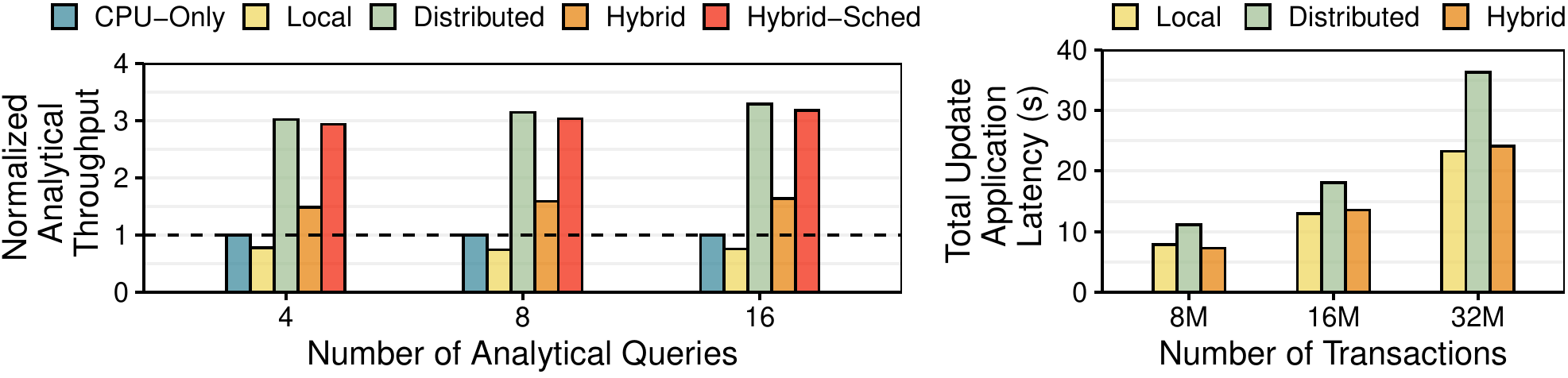}%
    \caption{\sgmod{Normalized analytical throughput (left) and update application latency (right) across different data placement and task scheduling strategies.}{Effect of data placement \gfcrii{and task} scheduling on \gfcri{analytical} throughput (left) and update application latency (right).}
    }
    \label{fig:eval-data-placement}
\end{figure}

\gfcri{Fig.}~\ref{fig:eval-data-placement} (left) shows the analytical throughput, \change{normalized to \gfcrii{a} \emph{CPU-\gfcri{O}nly} baseline \gfcrii{where one core services all queries to the same column,} for \gfcri{different number of analytical queries}}. We make four observations. 
First, \emph{Local} reduces throughput by 23.9\%\gfcri{,} on average\gfcri{,} over \emph{CPU-\gfcri{O}nly}, because in \emph{Local}, each analytical query can only use (1)~the PIM cores in the local vault, which cannot issue \gfcri{enough} memory requests concurrently \gfcri{to saturate the vault's memory bandwidth and exploit memory-level parallelism (MLP)~\gfcriii{\cite{mondrian,glew1998mlp, mutlu2003runahead, qureshi2006case, mutlu2008parallelism,mutlu2006efficient,mutlu2005techniques,chou2004microarchitecture,tuck2006scalable,cont-runahead,lee2009improving}}}, and 
(2)~a single vault's \gfcri{memory} bandwidth.
In contrast, \emph{CPU-\gfcri{O}nly} leverages out-of-order cores \gfcri{that can issue \gfcrii{many} memory requests in parallel \gfcrii{to multiple memory channels}, exploiting MLP \gfcrii{and higher memory bandwidth}.}
\gf{Second,} \gfcrii{\emph{Distributed}} improves \gfcrii{analytical} throughput by 4.1$\times$/3.1$\times$ over \emph{Local}/\emph{CPU-\gfcri{O}nly}\gfcri{, on average}. This is because under \gfcrii{\emph{Distributed}}, each column is partitioned across all vaults, allowing \gfcri{the analytical engines} to service each \gfcri{analytical} query using 
(1)~all PIM cores, and (2)~the entire internal \gfcri{memory} bandwidth of the \gfcri{3D-stacked} memory. However, \gfcrii{\emph{Distributed}} increases the update application latency, on average, by 45.8\% (\gfcri{Fig.}~\ref{fig:eval-data-placement}, right), and thus, degrades data freshness. This is because of the high update application costs (\gfcri{\cref{sec:proposal:analytic-engine:placement}}), which \emph{Local} does \emph{not} incur.
\gfcri{Third,} \emph{Hybrid} addresses the shortcomings of \emph{Local}, improving \gfcri{analytical} throughput by 57.2\% over \emph{CPU-\gfcri{O}nly}, while having a similar update application latency \change{(\SI{0.7}{\milli\second})}. This is because the local dictionary copies eliminate most of the remote accesses. However, the throughput under \emph{Hybrid} is 49.8\% lower than \gfcrii{\emph{Distributed}}, because each query is serviced only using resources (\gfcri{memory} bandwidth and \gfcri{PIM cores}) available in the local vault group\gfcrii{, leading to resource underutilization}. 
\gf{Fourth,} \emph{Hybrid-\sgcri{Sched}} overcomes \gfcrii{\emph{Hybrid}'s resource underutilization issues by enabling} task stealing, \gfcrii{which makes} idle resources in remote vaults available for analytical queries. \gfcrii{\emph{Hybrid-\sgcri{Sched}}} comes within 3.2\% of \gfcrii{\emph{Distributed}}, while maintaining the same update application latency as \emph{Hybrid}.

\gfcri{We conclude that our hybrid \gfcrii{data placement} along \gfcriii{with} our \gfcriii{optimized} \gfcrii{task} \gfcriii{scheduling} \gfcriii{heuristic} can provide high analytical throughput while fully leveraging the memory and computation resources of \titleShort's analytical island.}

\subsection{\gfcriv{Effect of the Dataset Size}}
\label{sec:eval:multiple}

\change{\gfcri{Fig.}~\ref{fig:eval-energy-multistack} (left) shows how \titleShort performs as the dataset size grows.
To accommodate the larger data, we increase the number of \gfcrii{memory} stacks, doubling the \gfcriii{dataset} size as we double the stack count. We use a workload with 32M transactional and 60K  analytical queries, and analyze analytical throughput normalized to \emph{Multiple-Instance}, as a case study. We assume stacks are connected together using a processor-centric topology~\gfcrii{\cite{conda, tom}}. To provide a fair comparison, we double the number of cores available to the analytical threads in the \emph{Multiple-Instance} baseline as we double the number of stacks, to compensate for the doubling of hardware resources available to \titleShort (since there
are twice as many vaults). 

\begin{figure}[ht]
    \centering
        \centering
        \includegraphics[width=\linewidth]{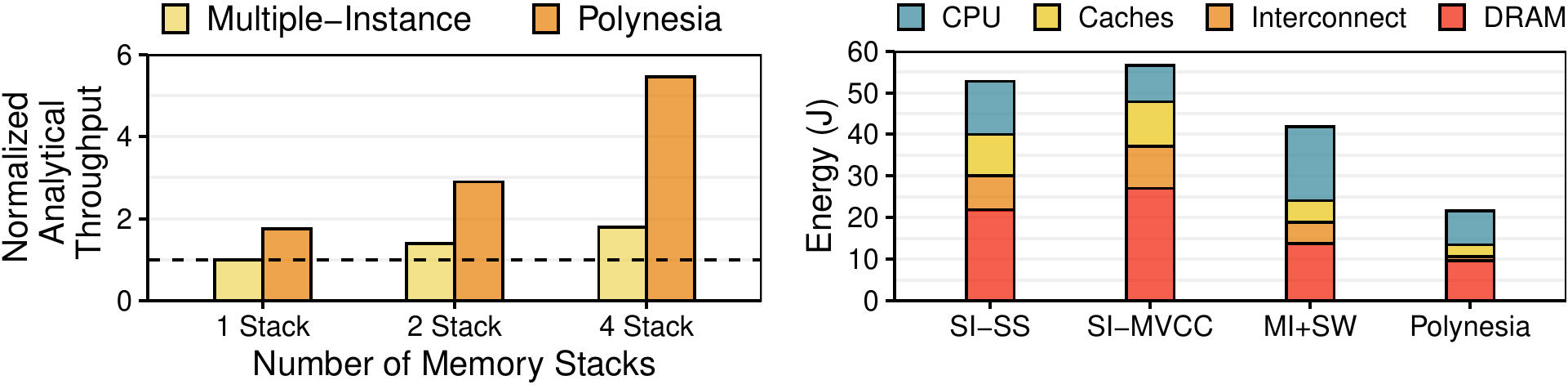}%
    \caption{\change{\gfcriv{Number of memory stacks vs.\ analytical throughput} (left). System energy (right).}}
    \label{fig:eval-energy-multistack}
\end{figure}

\gf{We make two observations. First,} \titleShort significantly outperforms \emph{Multiple-Instance} (up to 3.0$\times$) and scales well as we increase the stack count. This is because, as we increase \gfcrii{the} stack count, columns can be distributed more evenly across vault groups, which reduces the probability of multiple queries colliding in the same vault group. 
\gf{Second,} with increasing dataset size, the overheads of consistency mechanism, update propagation and analytical query execution are all higher for \emph{Multiple-Instance}, which hurts its scalability. The transactional throughput (not shown) decreases by 54.4\% at four stacks for \emph{Multiple-Instance}, compared to one stack, but decreases by \emph{only} 8.8\% for \titleShort.}

\gfcri{We conclude that \titleShort scales well with the \gfcrii{dataset size and thus the} number of 3D-stacked memories \gfcrii{used} in the HTAP system. We believe that further performance can be achieved by levering interconnection topologies that favor memory-centric architectures~\gfcrii{\cite{rezaei2020nom, tesseract,kim2013memory,dai2018graphh}}.}

\subsection{Energy Analysis}
\label{sec:eval:energy}
\gfrev{We model system energy
similar to prior work~\gfcriv{\cite{jeddeloh2012hybrid, google-pim, conda,amiraliphd}}, which sums the energy consumed
by the CPU cores, all caches (modeled using CACTI-P 6.5~\cite{CACTI}),
DRAM, and all on-chip and off-chip interconnects.} 
\gfcri{Fig.}~\ref{fig:eval-energy-multistack} (right) shows the total system energy across \gfcri{three} HTAP DBMSs. \gfcri{We make two observations.}
\gfcri{First, \titleShort consumes \gfcrii{only} 0.41$\times$/0.38$\times$/0.51$\times$ the energy of \emph{SI-SS}/\emph{SI-MVCC}/\emph{MI+SW}. \titleShort \emph{significantly} reduces energy consumption compared to the three HTAP DBMSs since it eliminates \gfcrii{a large fraction (30\%)} of off-chip accesses and uses custom logic and simple in-order PIM core\gfcri{s} for its analytical islands.
Second, \sgcri{the energy consumption of \emph{MI+SW}} is 0.8$\times$ and 0.7$\times$ that of \emph{SI-SS} and \emph{SI-MVCC}. However, \emph{MI+SW} still \gfcri{consumes more energy than \titleShort} since it (i)~cannot reduce the large number of memory accesses to off-chip memory, and (ii)~\gfcrii{uses} large and power-hungry CPU cores \gfcrii{and caches}. \info{We do not have energy results for MI+SW+HB}Such sources of energy consumption cannot be eliminated by simply providing high memory bandwidth to CPU cores.} \gfcri{We conclude that \titleShort is an energy-efficient HTAP DBMS.}

\subsection{Area Analysis}
\label{sec:eval:area}
\Copy{CR2/6}{\gfrev{We use previously reported data~\gfcrii{\cite{cortex-a7}} to determine the area of our PIM cores (used by the analytical islands), and Calypto Catapult~\gf{\cite{catapult}} to determine the area of our ASIC components, i.e, the update propagation unit and \gfcrii{the} copy unit (used by our consistency mechanism) for a \SI{22}{\nano\meter} process.}
\gfrev{\gfcrii{F}our PIM cores require
\SI{1.8}{\milli\meter\squared}, based on \gfcrii{ARM}
Cortex-A7 (\SI{0.45}{\milli\meter\squared} each)~\cite{cortex-a7}. The area Calypto Catapult reports for our ASIC components is \SI{0.7}{\milli\meter\squared} for the update propagation unit and \SI{0.2}{\milli\meter\squared} for the copy unit for our consistency mechanism. This
 brings \titleShort's total \gfcrii{hardware area} to \SI{2.7}{\milli\meter\squared} per vault.}
 
\gfrev{We conclude that our design can fit completely within the unused area in the logic layer of 3D-stacked memory (\SI{4.4}{\milli\meter\squared} per vault~\cite{google-pim,tetris,mondrian}). As a comparison, a 4-core Intel i5 processor with a 6 MB L3 cache has a die area of \SI{160}{\milli\meter\squared}~\cite{ivb-vs-snb}.} }

\section{Related Work}
\label{sec:related}

To our knowledge, this is the first work that 
(\gf{1)~proposes specialized hardware--software \gfcrii{co-designed} accelerators to cater for \gfcrii{heterogeneous workload demands in} HTAP systems, }
(\gf{2})~describes an HTAP system that meets all \gf{three} desired HTAP properties, \gf{and
(3)~uses processing-in-memory to alleviate data movement \gfcrii{overheads} in HTAP systems.} We briefly \gfcrii{summarize} related works.

\paratitle{HTAP Systems}
Several works from industry (e.g., \cite{sap-hana, oracle-dual-format, sql-htap,sap-soe, real-time-analysis-sql}) 
 and academia (e.g., \gf{\cite{hyper,peloton,hyrise,h2tap,l-store,batchdb,scyper, janus,janus-graph,sap-parallel-replication, sirin2021performance}})
 propose techniques to 
support HTAP. Many of them use a single-instance design~\cite{hyper, peloton,hyrise,h2tap,l-store,sap-hana}, while others are multiple-instance~\cite{batchdb,oracle-golden-gate,sap-soe}. All of these proposals suffer from the drawbacks we highlight
 in \S\ref{sec:bkgd:motiv}, and none can fully meet the \gfcrii{three} desired HTAP properties \gfcrii{(\cref{sec:bkgd:requirements})}.

\paratitle{Analytical Query Acceleration}
\gfcrii{Various} prior works focus solely on analytical 
workloads~\cite{mondrian,q100,JAFAR,widx,harp,Hash:NME}. Some of these works propose to use 
specialized on-chip accelerators~\cite{q100,widx,harp} while others propose to use PIM to 
speed up analytical operators~\cite{JAFAR,mondrian,Hash:NME}. However, none of these works study
the effect of data placement or task scheduling for the analytical workload in the context of PIM or HTAP systems.

\paratitle{\gf{Processing-in-Memory (PIM)}}
\gfcrii{Many} works~\gfcriv{\cite{google-pim,amiraliphd, graphp, tesseract,tetris,
data-reorganization-pim, graphpim,pim-graphics, pim-enabled,tom,
Mingyu:PACT, toppim, neurocube,guo2014wondp, PICA, cairo, pattnaik.pact16, cali2020genasm,syncron,fernandez2020natsa,boroumand2021google,kim.bmc18,NIM,tsai:micro:2018:ams,liu20173d, pugsley2014ndc, santos2018processing,gao2016hrl,rezaei2020nom,singh2019napel,boroumand2021google_arxiv,oliveira2021pimbench,lee2021hardware,kwon202125,balasubramonian2014near,nair2015active,zhuo2019graphq,huang2020heterogeneous,dai2018graphh,akin2014hamlet,loh2008stacked,zhu2013accelerating}}
add compute units to the logic layer of 3D-stacked memory \gfcrii{\gfcriii{to accelerate} various workloads}\gfdel{~\cite{hbm,hmcspec2,lee2016smla, kim.cal15}}. \sgmod{While these works propose various forms of in-memory hardware, none of them}{None of these works}
are designed for HTAP systems, and \sgdel{the works }are largely orthogonal. \gf{Prior PIM-based DBMS proposals~\cite{mondrian,Mingyu:PACT,Hash:NME,JAFAR, santos2017operand,tome2018hipe} solely focus on analytical workloads. \gfcrii{N}one of them study the effect of data placement or task scheduling for the analytical workload in the context of PIM or HTAP systems. \gfdel{As we discussed in Section~\ref{sec:proposal:analytic-engine} and showed in Sections~\ref{sec:eval:analysis} and \ref{sec:eval:analytical}, efficient analytical query execution strongly depends on (1) data layout and data placement, (2) the task scheduling policy, and (3) how each physical operator is executed. Like prior works~\cite{mondrian,Mingyu:PACT,q100}, we find that the execution of physical operators of analytical queries significantly benefit from PIM. However, we show that without an HTAP-aware and PIM-aware data placement strategy and task scheduler, PIM logic for operators alone cannot provide significant throughput improvements
One prominent example of prior PIM-based proposals for analytical workload is Mondrian~\cite{mondrian}. We evaluate the performance benefit of \titleShort’s analytical engine over Mondrian. Our analysis shows that Polynesia’s analytical engine performs on average 63.2\% better than Mondrian in terms of analytical throughput. The reason is that Mondrian accelerates only physical operators. Our scheduling policy and data placement enables us to efficiently execute analytical queries and achieve higher analytical throughput than Mondrian. Note that simply extending the Mondrian engine (or any prior PIM-based proposal for analytical workload) to support transactional workload does not address the key HTAP challenges, as we still need to provide data propagation and consistency mechanisms.}}

\section{Conclusion}

We propose \titleShort, a novel HTAP system that makes use of
\sgmod{multiple workload-optimized}{workload-optimized transactional and analytical} islands \gfcrii{(i.e., co-designed hardware/software units)} to enable real-time analytical queries
without sacrificing throughput. \sgdel{In \titleShort, transactional islands
make use of conventional transactional database engines running on
multicore CPUs, while analytical islands
make use of custom co-designed algorithms and hardware that are
placed inside memory.} Our analytical islands \sgdel{are designed to }alleviate
the data movement and workload interference \gfcrii{overheads} incurred in 
state-of-the-art HTAP systems, while ensuring that
data replicas for analytical workloads are kept up-to-date with
the most recent version of the transactional data replicas. \sgii{\titleShort outperforms three state-of-the-art HTAP systems
(with a 1.7$\times$/3.7$\times$ higher transactional/analytical throughput on average),
while consuming less energy (48\% lower than the best) \gfcrii{and meeting all three desired HTAP properties. We conclude that \titleShort is an \gfcriii{effective and} efficient architecture for HTAP systems. We hope that \titleShort \gfcriv{inspires future research and development in hardware/software co-designed HTAP systems that take advantage of processing-in-memory.}}}

\section*{Acknowledgments}
\gfcri{We thank the anonymous reviewers of MICRO 2019/2020, ASPLOS 2021, and ICDE 2022 for feedback. We thank SAFARI Research Group members for valuable feedback
and the stimulating intellectual environment they provide. We
acknowledge the generous gifts of our industrial partners, especially
Google, Huawei, Intel, Microsoft, and VMware. This research was
partially supported by the Semiconductor Research Corporation and \sgcri{the} ETH Future Computing Laboratory.}

{
\bibliographystyle{IEEEtran}
\bibliography{references}
}

\end{document}